\documentclass[twocolumn]{aastex6}

\usepackage{aas_macros}
\usepackage{amsfonts}
\usepackage{amsmath}
\usepackage{graphicx}
\usepackage{subfigure}

\newcommand{\fbol}{$F_{\mathrm{bol}}$}
\newcommand{\teff}{T$_{\mathrm{eff}}$}
\newcommand{\targ}{K2-264~}
\newcommand{\targtight}{K2-264} 
\newcommand{\vespa}{\texttt{vespa}}
\shorttitle{ZEIT VIII} 
\shortauthors{Rizzuto et al.}

\begin{document}
\title{Zodiacal Exoplanets in Time (ZEIT) VIII: A Two Planet System in Praesepe from K2 Campaign 16}

\author{
Aaron C. Rizzuto\altaffilmark{1,2,3},
Andrew Vanderburg\altaffilmark{1,4}, 
Andrew W. Mann\altaffilmark{5},
Adam L. Kraus\altaffilmark{1},
Courtney D. Dressing\altaffilmark{6},
Marcel A. Ag\"{u}eros\altaffilmark{7},
Stephanie T. Douglas\altaffilmark{8,9},
Daniel M. Krolikowski\altaffilmark{1}
}
\altaffiltext{1}{Department of Astronomy, The University of Texas at Austin, Austin, TX 78712, USA}
\altaffiltext{2}{arizz@astro.as.utexas.edu}
\altaffiltext{3}{51 Pegasi\,b Fellow}
\altaffiltext{4}{NASA Sagan Fellow}
\altaffiltext{5}{Department of Physics and Astronomy, The University of North Carolina at Chapel Hill, Chapel Hill, NC 27599, USA}
\altaffiltext{6}{Astronomy Department, University of California, Berkeley, CA 94720, USA}
\altaffiltext{7}{Department of Astronomy, Columbia University, 550 West 120th Street, New York, NY 10027, USA}
\altaffiltext{8}{Harvard-Smithsonian Center for Astrophysics, 60 Garden St, Cambridge, MA 02138}
\altaffiltext{9}{NSF Astronomy and Astrophysics Postdoctoral Fellow}

\begin{abstract}
Young planets offer a direct view of the formation and evolution processes that produced the diverse population of mature exoplanet systems known today. The repurposed \emph{Kepler} mission \emph{K2} is providing the first sample of young transiting planets by observing populations of stars in nearby, young clusters or stellar associations. We report the detection and confirmation of two planets transiting \targtight, an M2.5 dwarf in the 650\,Myr old Praesepe open cluster. Using our notch-filter search method on the \emph{K2} lightcurve, we identify planets with periods of 5.84\,d and 19.66\,d. This is currently the second known multi-transit system in open clusters younger than 1\,Gyr. The inner planet has a radius of 2.27$_{-0.16}^{+0.20}$\,R$_\oplus$ and the outer planet has a radius of 2.77$_{-0.18}^{+0.20}$\,R$_\oplus$. Both planets are likely mini-Neptunes. These planets are expected to produce radial velocity signals of 3.4 and 2.7\,m/s respectively, which is smaller than the expected stellar variability in the optical ($\simeq$30\,m/s), making mass measurements unlikely in the optical, but possible with future near-infrared spectrographs. We use an injection-recovery test to place robust limits on additional planets in the system, and find that planets larger than 2\,R$_\oplus$ with periods of 1-20\,d are unlikely.

\end{abstract}
\maketitle

\section{Introduction} 
Planets and their host stars can change dramatically over their lifetimes. Their structural, orbital, and atmospheric properties are all expected to evolve through interactions with their host star \citep[e.g.,][]{Ehrenreich2015}, the protoplanetary disk from which they formed \citep[e.g.,][]{Cloutier2013}, other planets in the system \citep[e.g.,][]{Chatterjee2008}, and the greater stellar environment \citep[e.g.,][]{Cai2018}. Understanding the underlying drivers and relative importance of these evolutionary mechanisms is critical for revealing the early sculpting of planetary systems and the conditions that give rise to the diversity of mature planetary systems revealed by {\it Kepler} and earlier exoplanet surveys \citep[e.g.,][]{2014Natur.513..336L,2016ApJ...817L..13I}.

Exoplanets likely evolve the most during their first Gyr \citep[e.g.,][]{Adams2006, Mann2010,Lopez2012}, and planets $<$1\,Gyr old are therefore powerful probes of the important drivers of exoplanet evolution. Fortuitously, the repurposed {\it Kepler} mission, {\it K2} \citep{Howell2014}, has surveyed a number of young clusters and star forming regions spanning $<10$Myr \citep[Taurus-Auriga;][]{Kraus2017a}, to $\simeq$650\,Myr \citep[Praesepe and Hyades;][]{Martin2018}, with Upper Scorpius \citep[$\simeq$10\,Myr;][]{Pecaut2012} and the Pleiades \citep[$\simeq$112\,Myr;][]{Dahm2015} spanning intermediate ages. 

The Zodiacal Exoplanets in Time (ZEIT) survey \citep{zeit1} was designed to identify and characterize transiting planets in these young clusters and star forming regions using the precise photometry from {\it K2} \citep{VanCleve2016}. The greater goal is to better understand how planets form and evolve by comparing the statistical properties of exoplanets of different ages and to older systems found during the original {\it Kepler} mission \citep{Borucki2010,Thompson2018}. Thus far we have identified planets in Hyades \citep{Vanderburg2018}, Praesepe \citep{zeit4}, and Upper Scorpius \citep{zeit3}, many of which were found near-simultaneously by similar surveys focusing on exoplanets in young stellar associations \citep[e.g.,][]{Obermeier2016, David2016b, Pepper2017, Ciardi2018, Livingston2018}. 

Multi-transiting planetary systems are uniquely useful for studying stellar and planetary properties. In cases where the planets' eccentricities can be independently constrained \citep[e.g., through dynamics;][]{Deck2016,Gillon2017}, multi-transiting systems can be used to constrain stellar densities with a precision that rivals eclipsing binaries \citep[e.g.,][]{mann17a}. Even with no information on the host star properties, differences between the measured transit duration of planets in the same system can be used to measure the relative eccentricities \citep{Kipping2012}. Multi-transit systems where planets undergo transit timing variations offer the best opportunity to measure the masses of small planets \citep[e.g.,][]{Deck2015,Hadden2017}. Lastly, these systems provide a measurement of the mutual inclination of planets, a probe of the entropy of a system and hence the role of dynamical disruptions from their (expected) initially flat configuration \citep[e.g.,][]{Figueira2012,Ballard2016}. 

Multi-transiting systems in clusters offer a unique route to study the dynamical properties of planets with known (young) ages. So far, there is only one known multi-transiting system in an open cluster: K2-136, a three-planet system in the 650\,Myr old Hyades cluster \citep{zeit6, Livingston2018,Ciardi2018}. Here, we present the discovery of two planets transiting the 650\,Myr old Praesepe cluster star \targ (JS 597; \citealt{jones91}) from its {\it K2} light curve. \targtight\ hosts two super-Earth/mini-Neptune-sized planets in short ($\approx$6 and $\approx$20 day) period orbits. We describe our discovery and follow-up observations in Section \ref{observationsanddatareduction}, and we describe our analysis to determine stellar properties in  Section \ref{sec:stelpars}. In Section \ref{limitsonadditionalplanets} we place limits on additional planets in the system, and in Sections \ref{transitfitting} and \ref{fpp} we describe our transit fitting to determine stellar parameters and our false positive probability analysis. Finally, in Section \ref{discussion}, we discuss the implications from discovering a second multi-transiting cluster system.

\section{Observations and Data Reduction}\label{observationsanddatareduction}

\subsection{K2 Observations and Transit Identification}
\emph{K2} observed \targ from 7 Dec 2017 to 25 Feb 2018 during Campaign 16. The raw pixel level photometry was calibrated using the \emph{Kepler} pipeline \citep{twicken10,stumpe12} prior to public release of the data on 30 May 2018. \targ was selected as part of four \emph{K2} guest observer programs in Campaign 16\footnote{GO16022 PI: Rebull, GO16033 PI: Gaudi, GO16052 PI: Stello, GO16060 PI: Ag\"{u}eros}, three of which selected \targ on the basis of membership in the Praesepe cluster.

\begin{figure*}
    \centering
    \includegraphics[width=\textwidth]{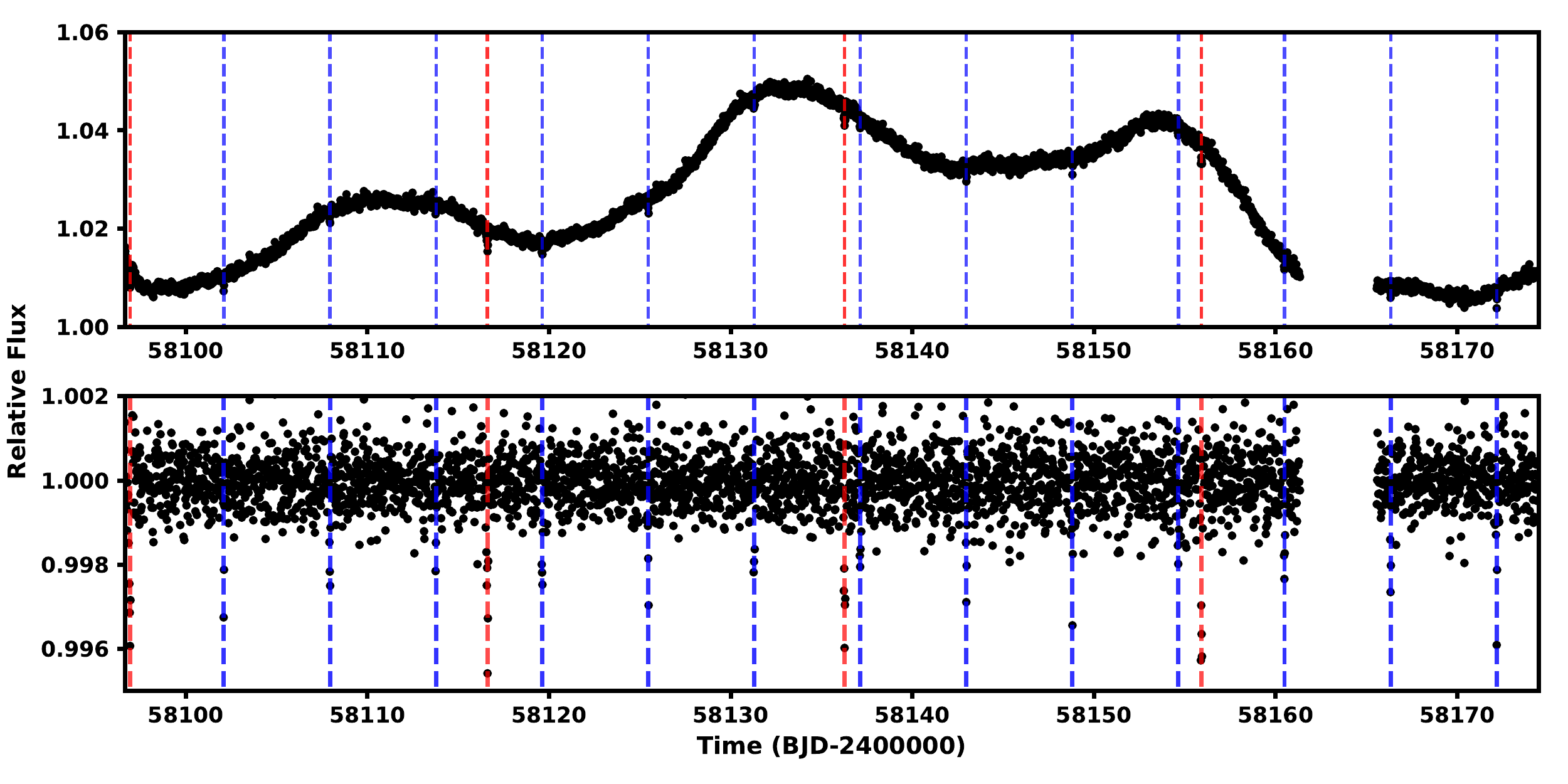}
    \caption{Lightcurve of \targ from \emph{K2} Campaign 16. The top panel shows the lightcurve after correction of the \emph{K2} roll/flat-field systematic following the method of \citet{vanderburgk2}. The lower panel shows the same lightcurve with the rotational variability removed while the transits were masked (post identification). Red and blue dashed lines indicate the outer (19.66\,day) and inner (5.84\,day) planets respectively.}
    \label{fig:lc}
\end{figure*}

We applied our detrending and transit search pipeline, which is described in detail in \citet{zeit5}, to the data for \targtight. To summarize the process, we first removed the \emph{K2} roll or flat-field systematic, caused by the instability of the \emph{K2} pointing and pixel-response variations using the method of \citet{vanderburgk2}. This produced a cleaned lightcurve that was mostly free of instrumental systematics but still contains signals from stellar variability and transiting planets. We removed the astrophysical variability with a ``notch-filter,'' which fits a 1-day window of the light curve as a combination of an outlier-resistant second-order polynomial and a trapezoidal notch. The inclusion of the notch allows aggressive detrending outside the notch without over-fitting that may remove or weaken transit-like signals. This window is then moved along each point in the light-curve, detrending the variability signal from the entire dataset. The periodic transits were then identified using the Box Least Squares algorithm \citep{kovacs02} on the detrended lightcurve. Figure \ref{fig:lc} shows the rotational variability, detrended lightcurve, and detected transit signals. 

Once the two transiting planets were detected, we re-extracted the data using a simultaneous fit to the \emph{K2} roll systematic, low-frequency stellar variability, and transits, including outlier rejection as described in \citet{vanderburg16}. The final lightcurve, following flattening by removal of the best-fit low-frequency variability and significant outliers, was then used for our MCMC transit fitting described below.

\subsection{NIR Spectra from SPEX}

\begin{figure}
    \centering
    \includegraphics[width=0.49\textwidth]{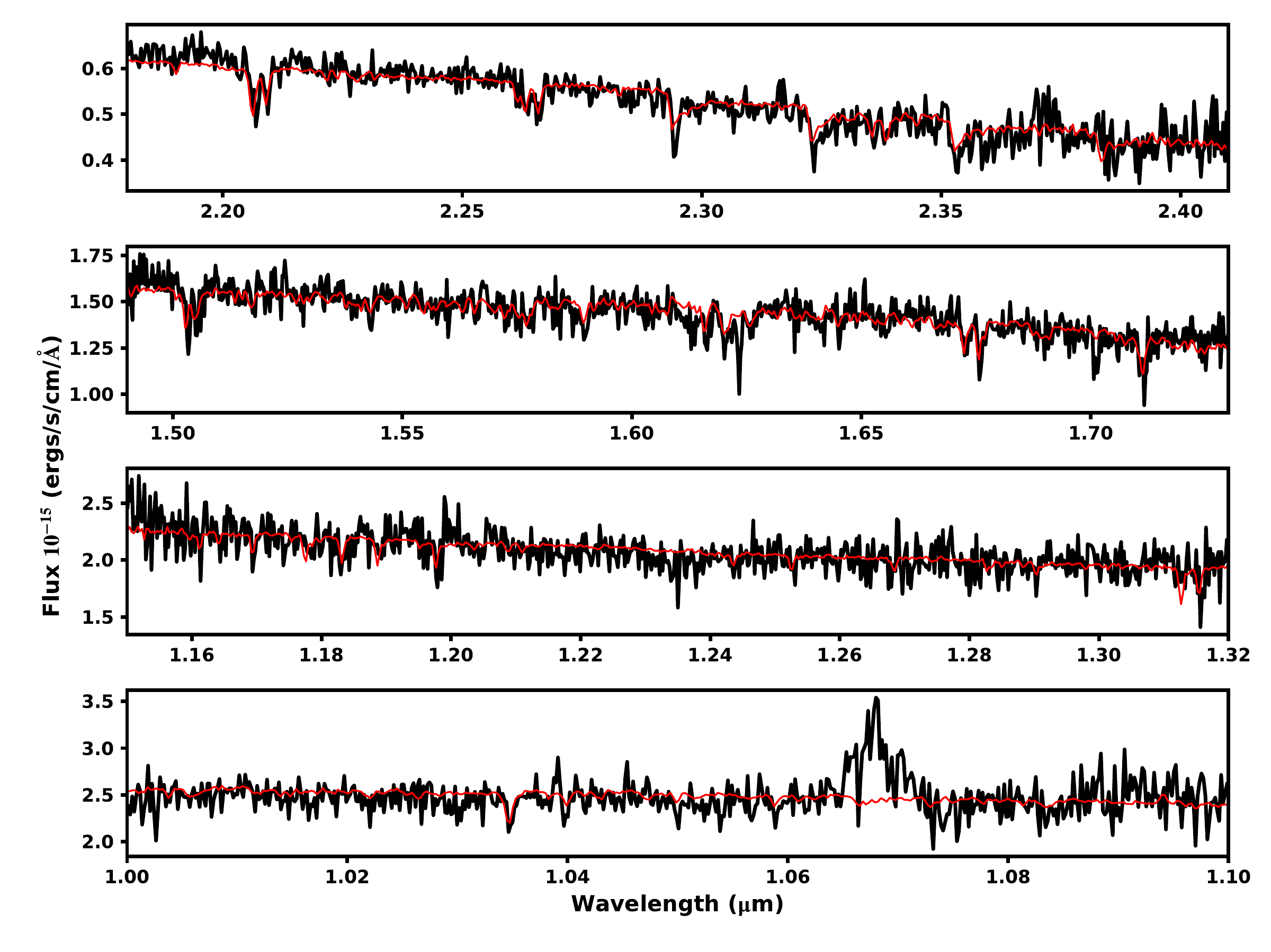}
    \caption{Spex NIR spectrum orders of \targtight. The black line shows the reduced data for \targ, and the red line is an M3 template spectrum for J1142+2642 taken with the same instrument.} 
    \label{fig:spex}
\end{figure}

On 2 June 2018 we observed \targ with the InfraRed Telescope Facility (IRTF) SpeX medium-resolution spectrograph \citep{rayner_et_al2003, rayner_et_al2004}. We used the 0.3\arcsec\ slit in SXD mode, which yielded a spectral resolution of $R \simeq2000$ from 0.7 to 2.55\,$\mu$m. Extraction and calibration of the spectrum, including flat, bias, and wavelength correction, was carried out using the Spextool package \citep{spextool}, which incorporates the xtellcor package \citep{vacca2003} to correct for telluric contamination. The observation was taken in poor conditions at high airmass and had a median signal-to-noise ratio (SNR) per pixel of $\simeq$15 in the first two orders covering 1.4-2.5\,$\mu$m, and SNR of $\simeq$10 in the two orders covering 0.9-1.3\,$\mu$m. Given the low SNR of the spectrum, we did not attempt to extract stellar properties such as T$_{\mathrm{eff}}$, $\log{g}$, or metallicity. Figure \ref{fig:spex} displays the resulting spectrum of \targtight.

We measured a radial velocity from the spectrum by cross-correlating with a similar spectral-type standard. This was done over each order using the \emph{tellrv} package\footnote{https://github.com/ernewton/tellrv} \citep{newton_tellrv}. After correcting for Barycentric motion, the cross-correlation yielded a radial velocity for \targ of 26$\pm$6\,km/s, which is within $\sim$1-$\sigma$ of the expected expected radial velocity for a Praesepe member.

\subsection{Literature Photometry and Astrometry}\label{sec:phot}
Photometry from multiple all-sky surveys were compiled to build a full SED for \targ. Optical \emph{g\,r\,i\,z} magnitudes were taken from the PanSTARRS point source catalog \citep{panstarrs_cat}. Near-IR {\emph J,H} and \emph{Ks} photometry was taken from the Two Micron All Sky Survey (2MASS; \citealt{2mass}), the \emph{r'} magnitude was taken from the Carlsberg Meridian Catalog (CMC15; \citealt{cmc15}). Mid-IR magnitudes in the W1-4 bands were taken from the Wide-Field Infrared Survey Explorer (WISE; \citealt{wise10}). \targ was not detected in the W3 and W4 bands ($\sim$12 and 24\,$\mu$m) and only upper limits were provided, so we excluded them from our analysis. Proper motions, parallax, and $G$, $RP$, and $BP$ magnitudes were taken from the \emph{Gaia} mission second data release \citep{gaiadr2_arxiv}. These data for \targ are shown in Table \ref{proptab}.

\subsection{Archival Imaging}

We examined archival imaging observations of \targ from several different surveys to search for nearby stars that might contribute the transit signals we see. In particular, we examined observations from the Palomar Observatory Sky Survey (POSS-I) to identify background stars at the present-day position of \targ, and we used observations from the Pan-STARRS survey to search for nearby faint companions. 

\begin{figure*}
    \centering
    \includegraphics[width=\textwidth]{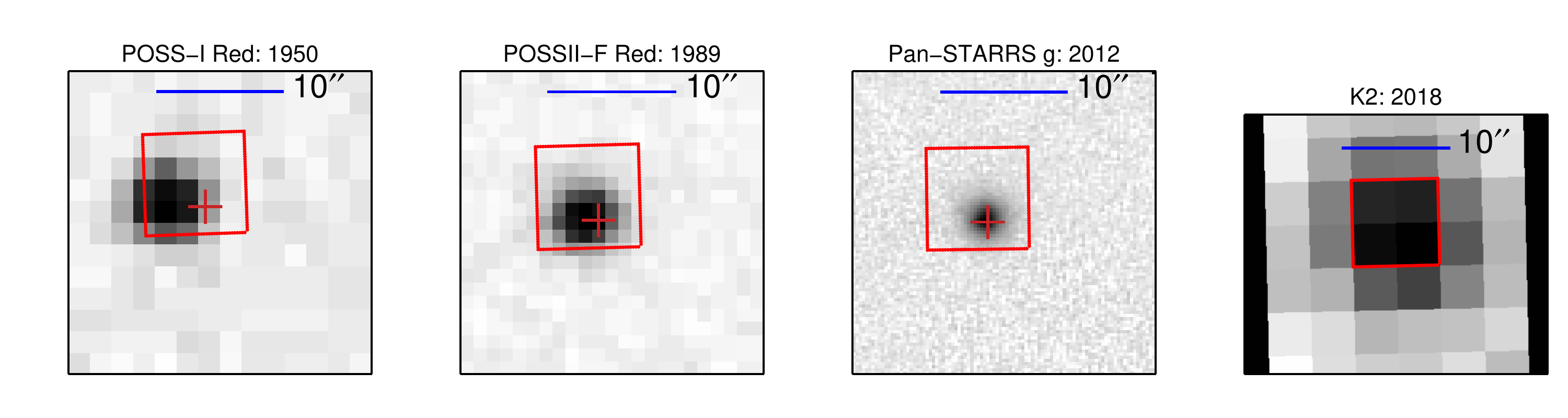}
    \caption{Archival imaging of \targtight, the red cross indicates the present-day position of the target taken from Gaia DR2 \citep{gaiadr2_arxiv}, and the red box indicates the \emph{K2} pixels used to make the lightcurve. \emph{Far left}: A POSS-I image  taken in 1950 of \targ on a photographic plate with a red-sensitive emulsion. The star's proper motion has caused its apparent position to move by several arcseconds since 1950, excluding the presence of background contaminants at the present day position of \targ. \emph{Middle left}: A POSS-II image taken in 1989 on a photographic plate with a red-sensitive emulsion. \emph{Middle right}: An image taken by the Pan-STARRS survey in g-band. The higher resolution and deeper Pan-STARRS images show no signs of nearby stars that might contribute the transit signals we detect toward \targ. \emph{Far right}: Summed \emph{K2} image of \targ.}
    \label{fig:archivalimaging}
\end{figure*}

We first used the POSS images of \targ to rule out the presence of bright background stars at the present-day position of the target star. \targ was observed by POSS in 1950, when its position was $\sim$2.7\arcsec\, away from its present day position due to the star's proper motion ($\mu = 40.1 \pm 0.1$ mas yr$^{-1}$). \targtight's PSF partially overlaps its present-day position, but it is still possible to rule out some  nearby companions. Based on nearby stars observed at the same time, we estimate that if there was a background star at the present-day position of \targ brighter than $R \sim 19$ mag, we would be able to detect it. Since we see no evidence for such a star, we can rule out the presence of these background companions about three magnitudes fainter than \targtight. We show the POSS image in Figure \ref{fig:archivalimaging}.

We also used observations from the Pan-STARRS survey to search for and rule out faint stars near the position of \targtight. Neither a query of the Pan-STARRS archive point source catalog nor visual inspection of images revealed any nearby stars closer to \targ than 30". With Pan-STARRS, we can rule out nearby stars to fairly faint limits (r $\gtrsim$ 20).  The Pan-STARRS image is shown in Figure \ref{fig:archivalimaging}.

\subsection{Companion Constraints from \emph{Gaia} Data Release 2}
\label{gaiaastro}
While detection limits for additional sources surrounding stars in the \emph{Gaia} second data release \citep{gaiadr2_arxiv} are not characterized by the \emph{Gaia} team, limits can be estimated using populations of known binaries detected in ground-based imaging surveys. \citet{zeigler18} used a sample of 620 binary companions to \emph{Kepler} Objects of Interest (KOIs) detected with Robo-AO imaging to characterize the detectability of companions 1-4" from a primary in the \emph{Gaia} second data release. \citet{zeigler18} find that companions with separations $<$1 arcsecond are not listed as separate sources in the \emph{Gaia} catalog, and provide contrast limits out to separations of 4". 

This method can be extended to smaller separations by examining the quality of the \emph{Gaia} astrometric fit, in particular the significance of the ``extra-error'' term. In order to do this we supplemented 363 high-confidence binary companions to KOIs identified by Robo-AO \citep{robokoi1,robokoi2,robokoi3,robokoi4} with 93 companions detected at $\rho < 1$\arcsec\, using imaging or aperture mask interferometry with the Near Infrared Camera (NIRC2) 
on the Keck 2 telescope by \citet{koi_binary_paper}. The higher spatial resolution of Keck, particularly when combined with aperture masking, provided companions down to $\rho \simeq 10$--20\,mas. The Robo-AO LP600 filter is very similar to the \emph{Gaia} \emph{G} bandpass \citep{zeigler18}, however the companions from \citet{koi_binary_paper} were detected in K-band. Under the assumption that these companions were very likely to be bound due to the small separations, we interpolated \emph{Gaia} \emph{G} band primary-secondary contrasts using the K-band contrast, the primary estimated effective temperature from \citet{koi_binary_paper}, and a 2\,Gyr PARSEC 1.2s isochrone \citep{chen14_padova}.

We then queried the \emph{Gaia} second data release (DR2) catalog in a 10 arcsecond cone around each KOI with a detected companion. We assessed detection by \emph{Gaia} on the basis of three separate conditions: (1) The companion was identified as a unique source in the catalog at the expected position angle and separation and with the expected contrast (2) The companion was not resolved as a distinct source in the \emph{Gaia} catalog, but the astrometric extra error significance ($D$) was $>$10-$\sigma$. This was only used for companions with separations $<$1 arcsecond. (3) The primary star was missing from the \emph{Gaia} catalog, again this condition was only applied to companions with separations of $\rho < 1$\arcsec. Our interpretation assumes that clear binaries where the astrometric solution was extremely poor were removed from the second data release, which is listed in \citet{gaiadr2_arxiv} as the intended operating procedure employed by the \emph{Gaia} data reduction team. Finally, if the contrast of the companion and the magnitude of the primary would indicate a \emph{Gaia} \emph{G} magnitude of the secondary of $>$21\,mag, we removed that companion from the test sample as it falls below the faint limit for the \emph{Gaia} survey and may not be robustly detected.

\begin{figure}
    \centering
    \includegraphics[width=0.49\textwidth]{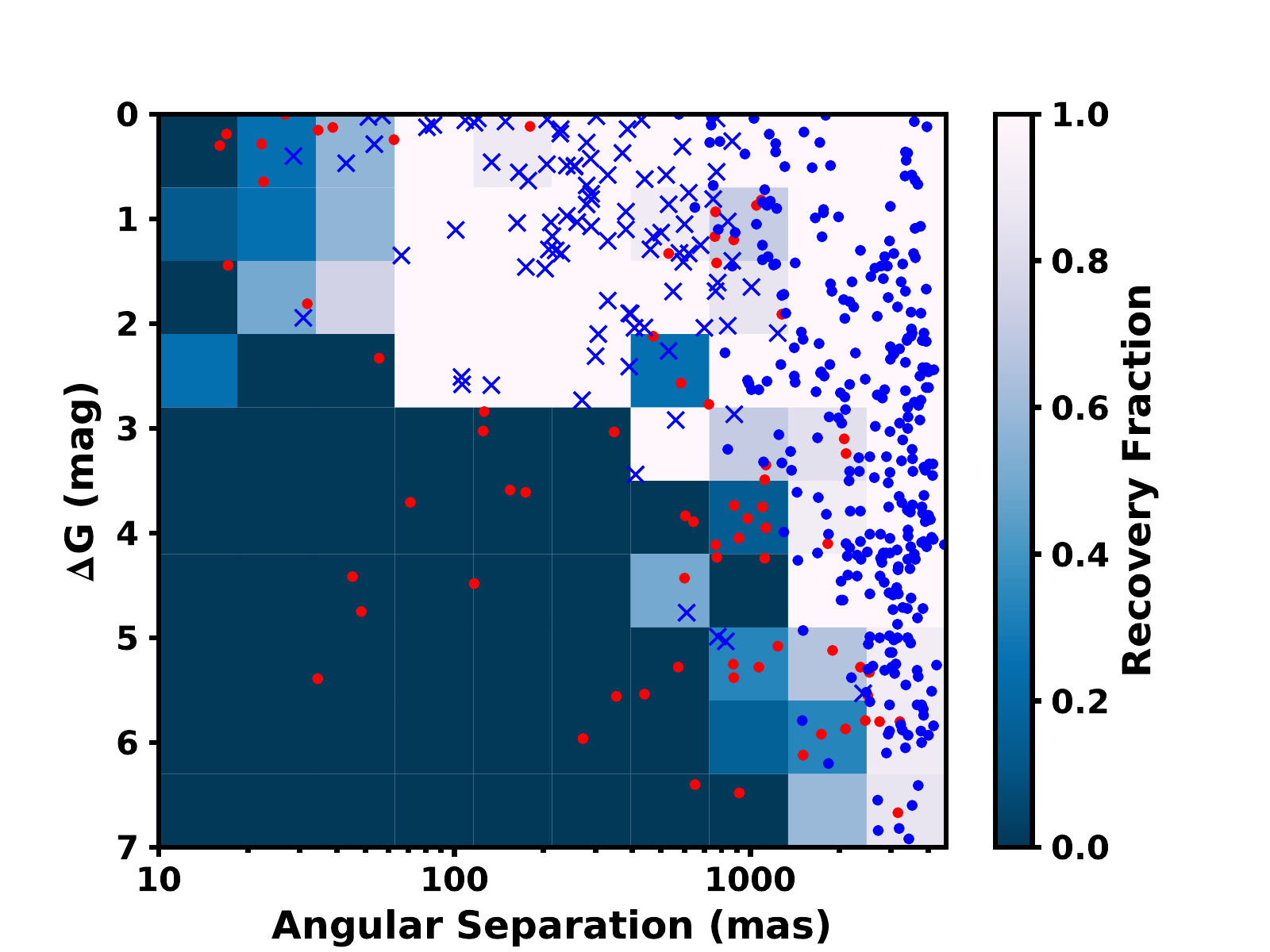}
    \caption{Recovered (blue) and missed (red) known companions to 457 Kepler objects of interest from \citet{robokoi1,robokoi2,robokoi3,robokoi4} and \citet{koi_binary_paper}. Red points indicate companions not recovered by \emph{Gaia}, blue crosses indicate astrometrically recovered companions, and blue points indicate companions resolved as separate sources in the \emph{Gaia} DR2 source catalog.}
    \label{fig:gaia_bindet}
\end{figure}

Figure \ref{fig:gaia_bindet} displays the separation and contrast of the recovered and non-recovered companions in the \emph{Gaia} second data release. We find similar magnitude limits in the 1-3\,arcsecond range as \citet{zeigler18}, with 50\% recovery for $\Delta$\emph{G} = 3\,mag at 1 arcsecond and for $\Delta$G = 6\,mag at 3". Inside 1\,arcsecond, companions with $\Delta$\emph{G} $<$ 2\,mag are reliably detected on the basis of the astrometric fit down to separations of 80\,mas. 

There were no sources within 35 arcsecond of \targ in \emph{Gaia} DR2, and the astrometric extra error significance for \targ is $D=4.98$-$\sigma$. The \emph{Gaia} DR2 release notes \citep{gaiadr2_arxiv} state that for stars with well behaved astrometry, $D$ should be considered as a half-normal with mean zero and spread of unity. Furthermore, \emph{Gaia} DR2 astrometry may contain instrument and attitude modelling errors that may inflate the value of $D$ \citep{gaiadr2_arxiv}. A value of $D=4.98$ is thus not anomalously large considering the number of sources in the Gaia catalog. Hence, we can rule out companions with contrasts of less than 2\,mag at separations of 80-1000\,mas.

\section{Stellar Parameters}
\label{sec:stelpars}

\begin{figure}
    \centering
    \includegraphics[width=0.49\textwidth]{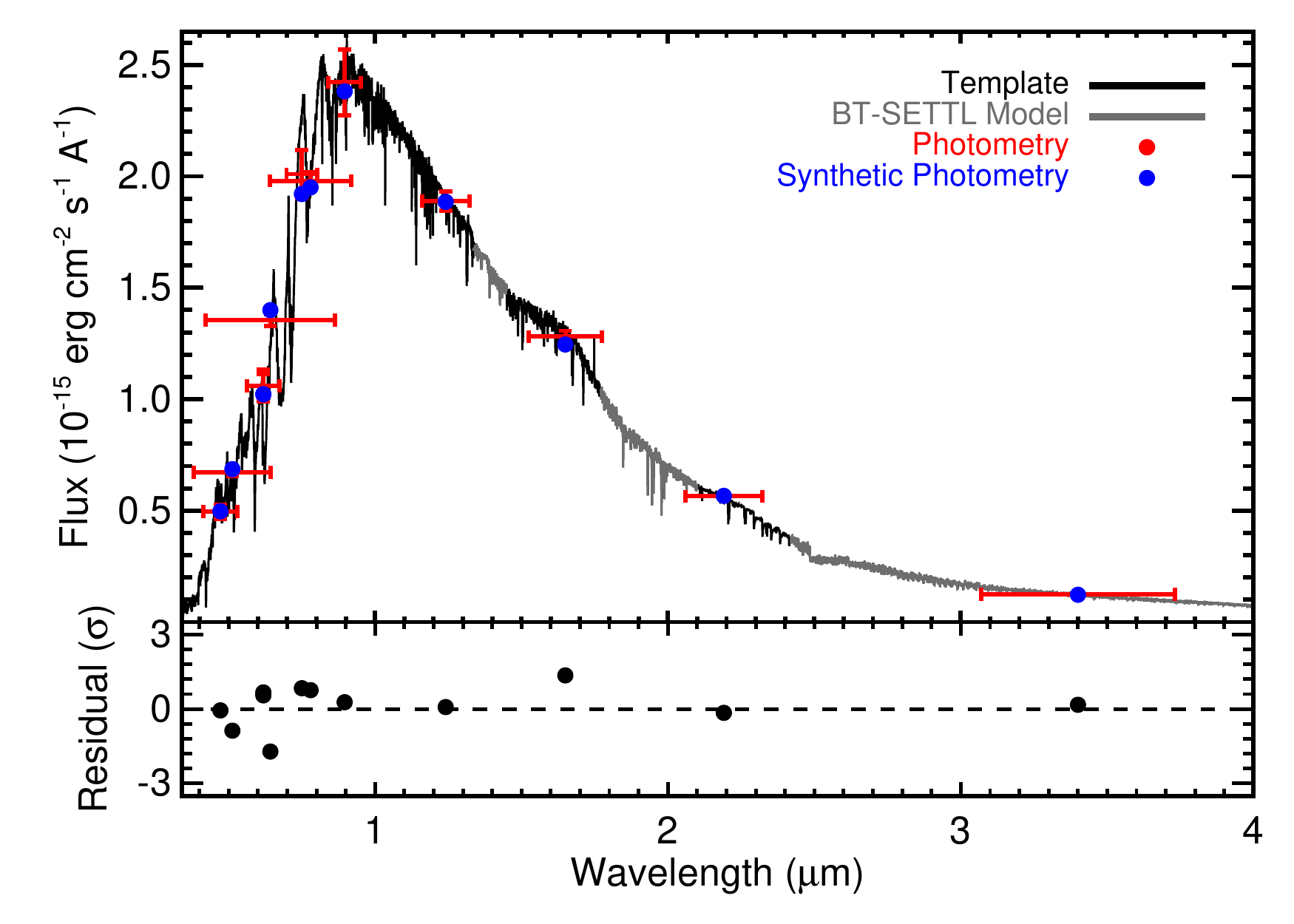}
    \caption{Best-fit spectral template compared to the photometry of \targtight.The template is constructed from observations (black) from \citet{Gaidos14} with a BT-SETTL model atmosphere of the same temperature interpolated into missing regions of the spectrum (gray). Measured photometry is shown in red, with vertical errors corresponding to the uncertainty in the flux and horizontal error bars showing the FWHM of the filter profile. Blue points mark the synthetic flux measurements derived by convolving the spectrum with the relevant filter profile. The bottom plot shows the difference between the photometry and synthetic photometry in units of standard deviations.}
    \label{fig:sed}
\end{figure}

\emph{Effective Temperature and Bolometric Flux:} We simultaneously solved for the spectral type and bolometric flux (\fbol) by fitting the literature photometry (Section~\ref{sec:phot}) using a grid of M-dwarf templates, following the technique outlined in the previous papers in this series (e.g., \citealt{zeit4,zeit6}). For the templates, we used a set of flux-calibrated templates of members of the Hyades open cluster, which were observed as part of programs to characterize nearby M dwarfs \citep{Gaidos14}.  We first filled missing regions of the  template spectra for which data were not available with BT-SETTL atmosphere models \citep{btsettl} of the corresponding temperature, and then reddened each template according to the E(B-V) value for Praesepe from \citet{Taylor06}. For each template, we computed synthetic magnitudes using the filter profiles and zero-points from \citet{DR2_photcal}\footnote{also see \href{https://gea.esac.esa.int/archive/documentation/GDR1/Data_processing/chap_cu5phot/sec_phot_calibr.html}{\emph{Gaia} photometric calibration documentation}}, \citet{mann15m} for other optical bands, and \citet{cohen03} for 2MASS. We compared these synthetic magnitudes to the archival values, letting the template choice and overall flux levels shift as free parameters. For each template, we computed \fbol\ by integrating over the full spectrum. Our final adopted spectral type and \fbol\ were those corresponding to the best-fit template weighted by the $\chi^2$ values from the comparison between observed and synthetic (from the templates) photometry). This method yielded a spectral type of M2.5($\pm$0.5) and and \fbol\ of $3.068\pm0.068\times10^{-11}$ erg\,cm$^{-2}$\,s$^{-1}$. The errors in \fbol\ account for variations due to many possible template fits, uncertainties in the cluster reddening, and uncertainties arising from interpolating over gaps in the spectrum. We show the best-fit template and a comparison to the photometry in Figure~\ref{fig:sed}.

To determine $R_*$, $M_*$, and $\rho_*$ we used the empirical $M_{K_S}$-$R_*$ relation from \citet{mann15m} and the $M_{K_S}$-$M_*$ relation from Mann et al. (submitted)\footnote{\href{https://github.com/awmann/M_-M_K-}{https://github.com/awmann/M\_-M\_K-}}. We computed $M_{K_S}$ from the \emph{Gaia} distance and 2MASS $K_S$. This yielded a radius of $0.473\pm0.014$\,R$_\odot$, a mass of $0.471\pm0.012$\,M$_\odot$, and a density of $4.46\pm0.40$\,$\rho_\odot$.  We can also assess the accuracy of the radius derived from the $M_{K_S}$-$R_*$ relation of \citet{mann15m} using the Stefan-Boltzmann equation, our bolometric flux, temperature of the best-fit template star, and the \emph{Gaia} parallax. We find that the radius corresponding to the best-fit temperature is 0.475$\pm$0.018\,R$_\odot$, which agrees with the radius from the $M_{K_S}$-$R_*$ very closely.

To calculate the total stellar luminosity, we combined our \fbol\ value with the \emph{Gaia} parallax, which yielded $0.0330\pm0.0012L_\odot$. Joining this with our radius determination and the Stefan-Boltzmann equation produced a \teff\ of 3580$\pm70$\,K. This \teff\ value was consistent with the assigned value for our best-fit template (3560$\pm$60\,K) derived by comparison to BT-SETTL models \citep{btsettl}, as described in \citet{Mann2013c}.  

\begin{figure}
    \centering
    \includegraphics[width=0.47\textwidth]{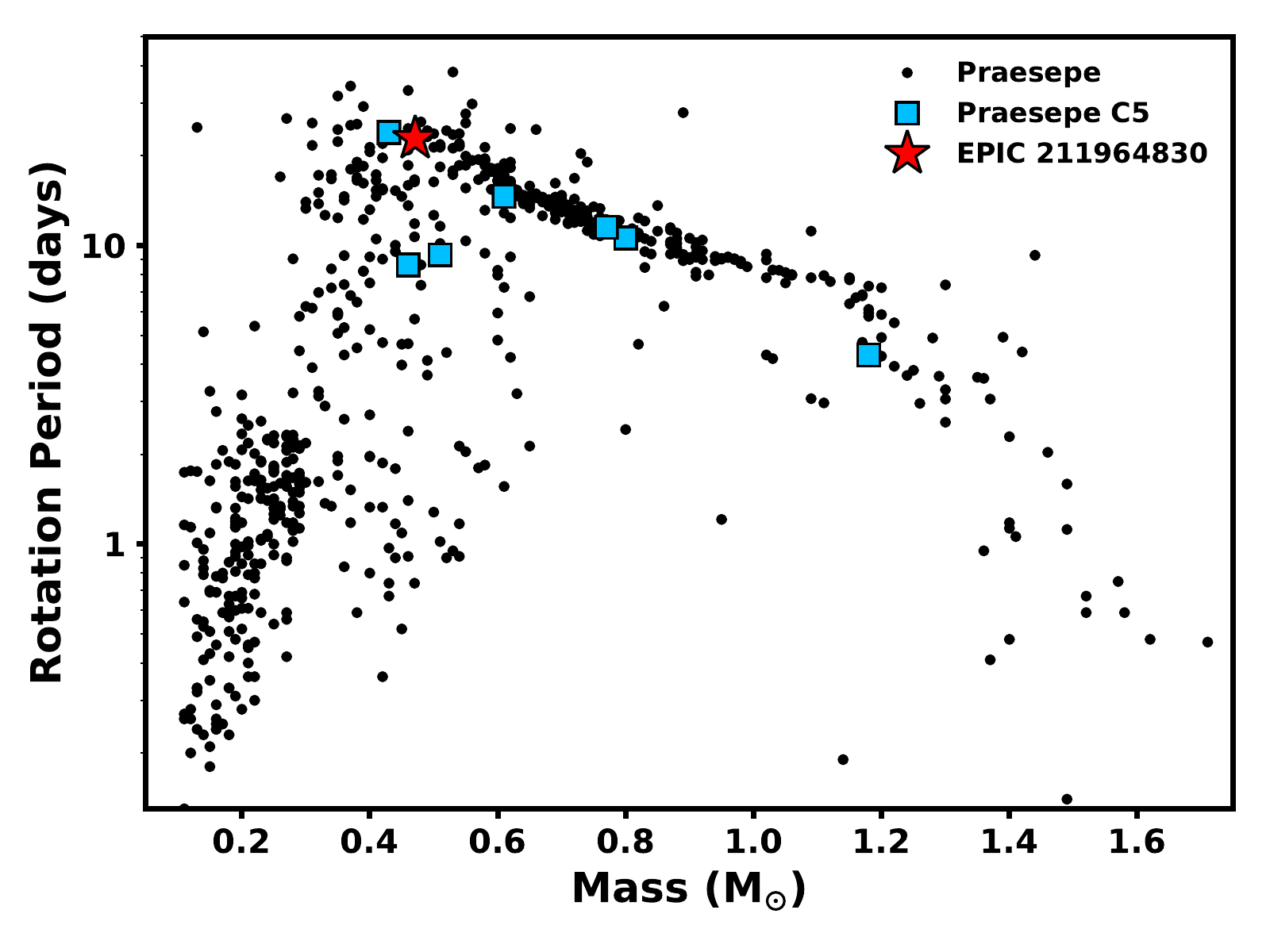}
    \includegraphics[width=0.48\textwidth]{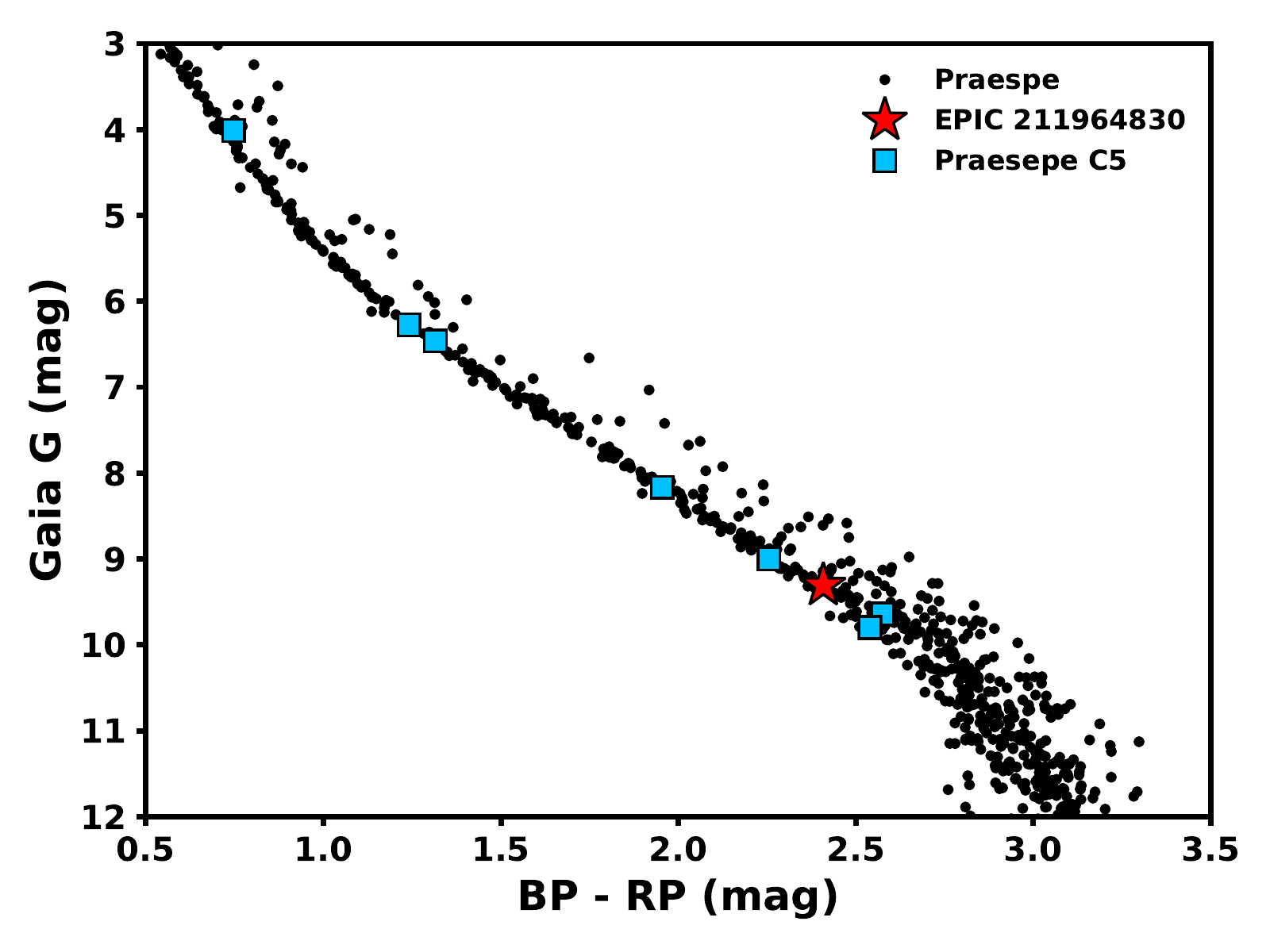}
    \caption{\emph{Top:} Rotation periods as a function of estimated stellar mass for Praesepe members from \citet{douglas17}. The red star indicates the rotation period of \targ which closely matches the Praesepe sequence. \emph{Bottom:} Color-magnitude diagram using \emph{Gaia} DR2 \emph{G, RP} and \emph{BP} magnitudes and parallax of Praesepe members from \citet{kraushillenbrand_comaber} with membership probabilities greater than 95\%. \targ (red star) lies on the tight single-star sequence of cluster members. In both panels the blue squares are the host stars of the other seven transiting planets in Praesepe identified in \emph{K2} C5 \citep{zeit4}.}
    \label{fig:prot_praesepe} 
\end{figure}

\emph{Rotation Period:} To determine the rotation period, we took the \emph{K2} roll-corrected light curve prior to removing the stellar variability, masked out the transits from the data, and computed a Lomb-Scargle periodogram spanning periods of 1--40 days. We fit a Gaussian to the largest peak in the periodogram to find the period at the peak power, and conservatively estimate the uncertainty as the standard deviation of the Gaussian divided by the periodogram power. We find the rotation period to be 22.8$\pm$0.6\,days. The rotation period of \targ lies directly on the Praesepe rotation-mass sequence. Figure \ref{fig:prot_praesepe} shows the rotation period of \targ in relation to the host stars of the seven other known Praesepe members with transiting planets \citep{zeit4} from \emph{K2} Campaign 5, and the full Praesepe population \citep{douglas17}.

\begin{figure}
    \centering
    \includegraphics[width=0.48\textwidth]{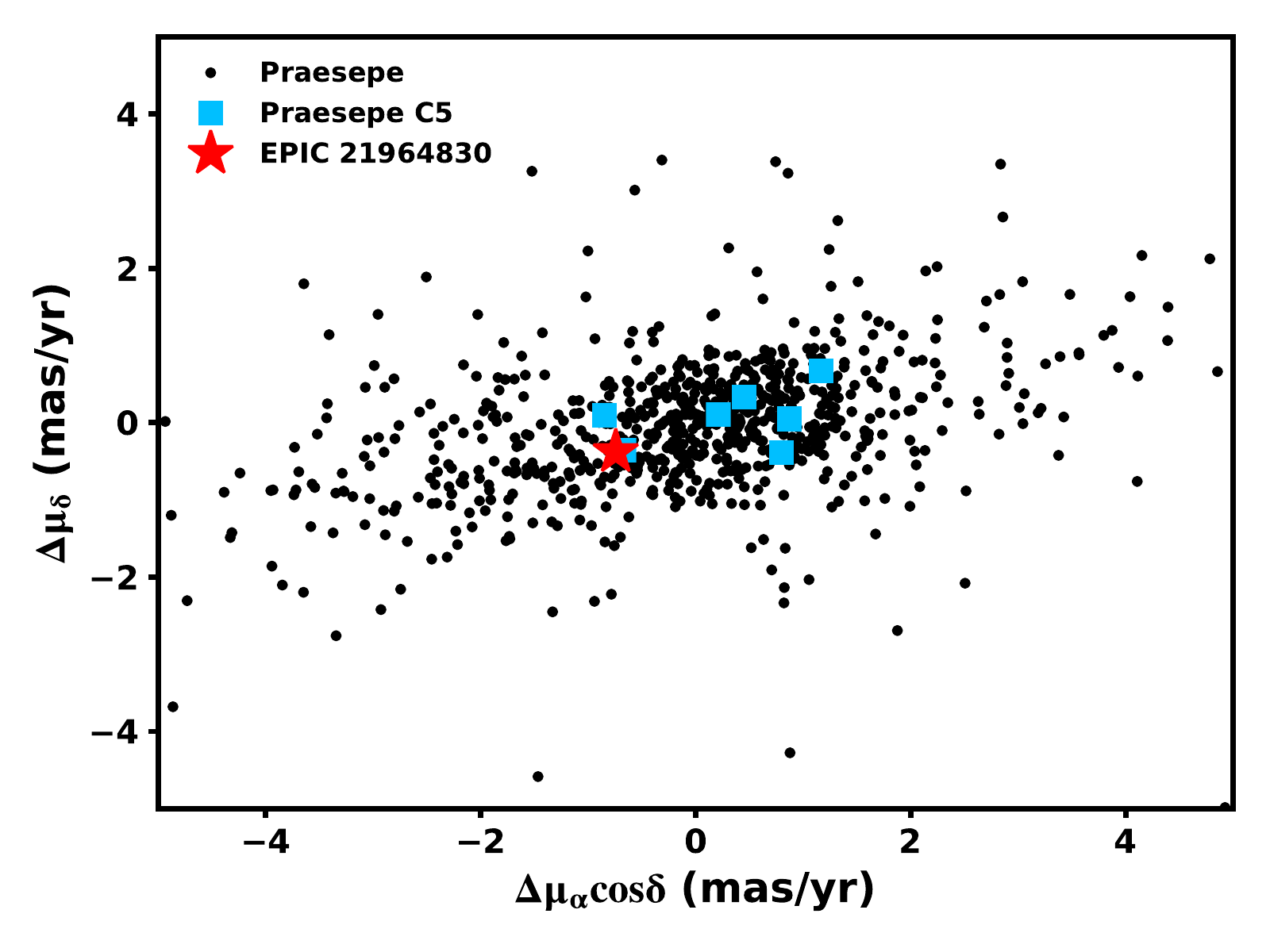}
    \caption{Proper motions offset from the expected cluster motion for \targ compared to known Praesepe members from \citet{kraushillenbrand_comaber} with membership probability greater than 95\%. The intra-cluster dispersion appears to be 1-2\,mas/yr, or equivalently 1-2\,km/s. The proper motion of \targ is highly consistent with the projected sky motion of the Praesepe cluster.}
    \label{fig:pmproj}
\end{figure}

\begin{figure*}
    \centering
    \includegraphics[width=0.3\textwidth]{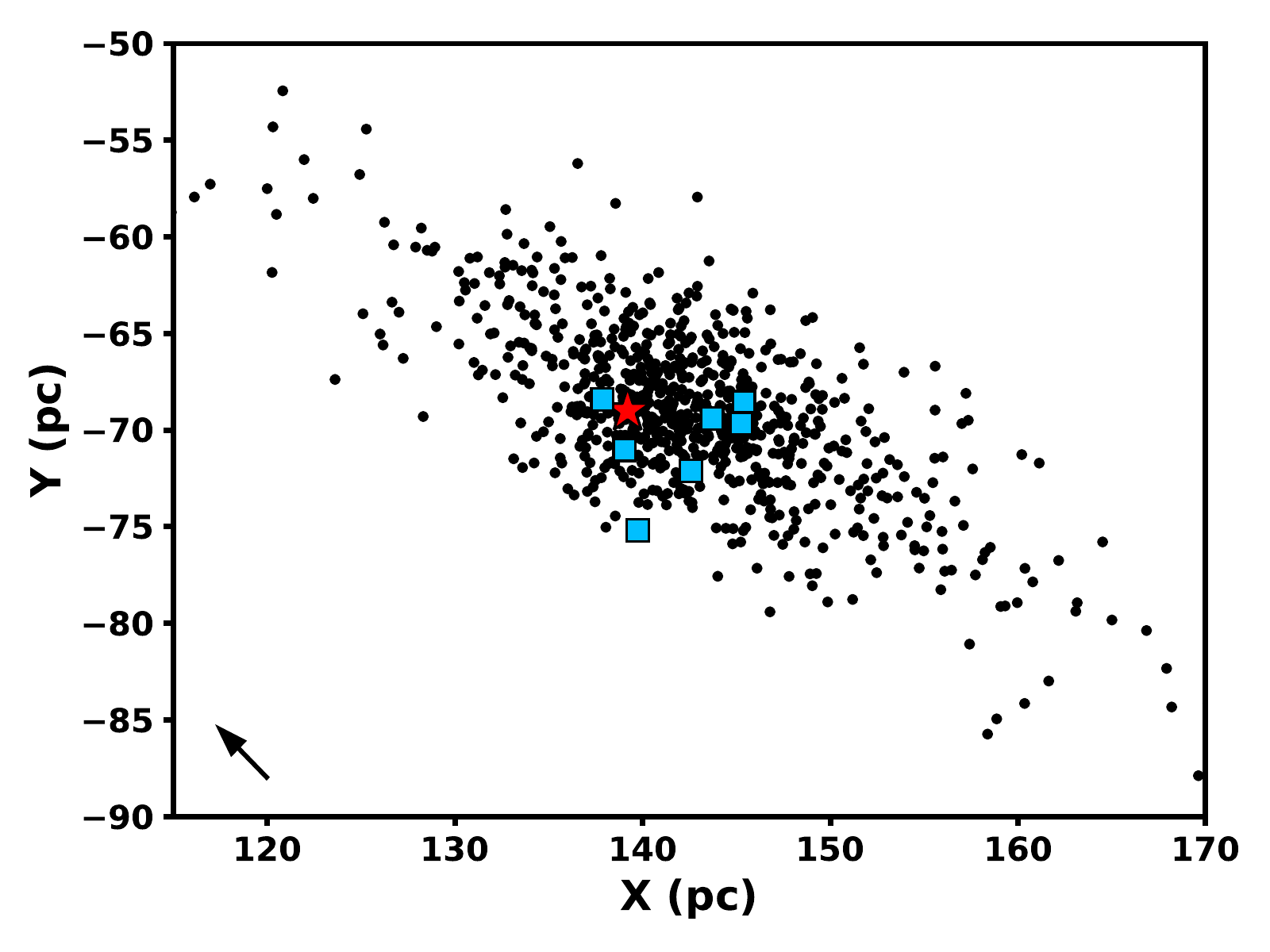}
    \includegraphics[width=0.3\textwidth]{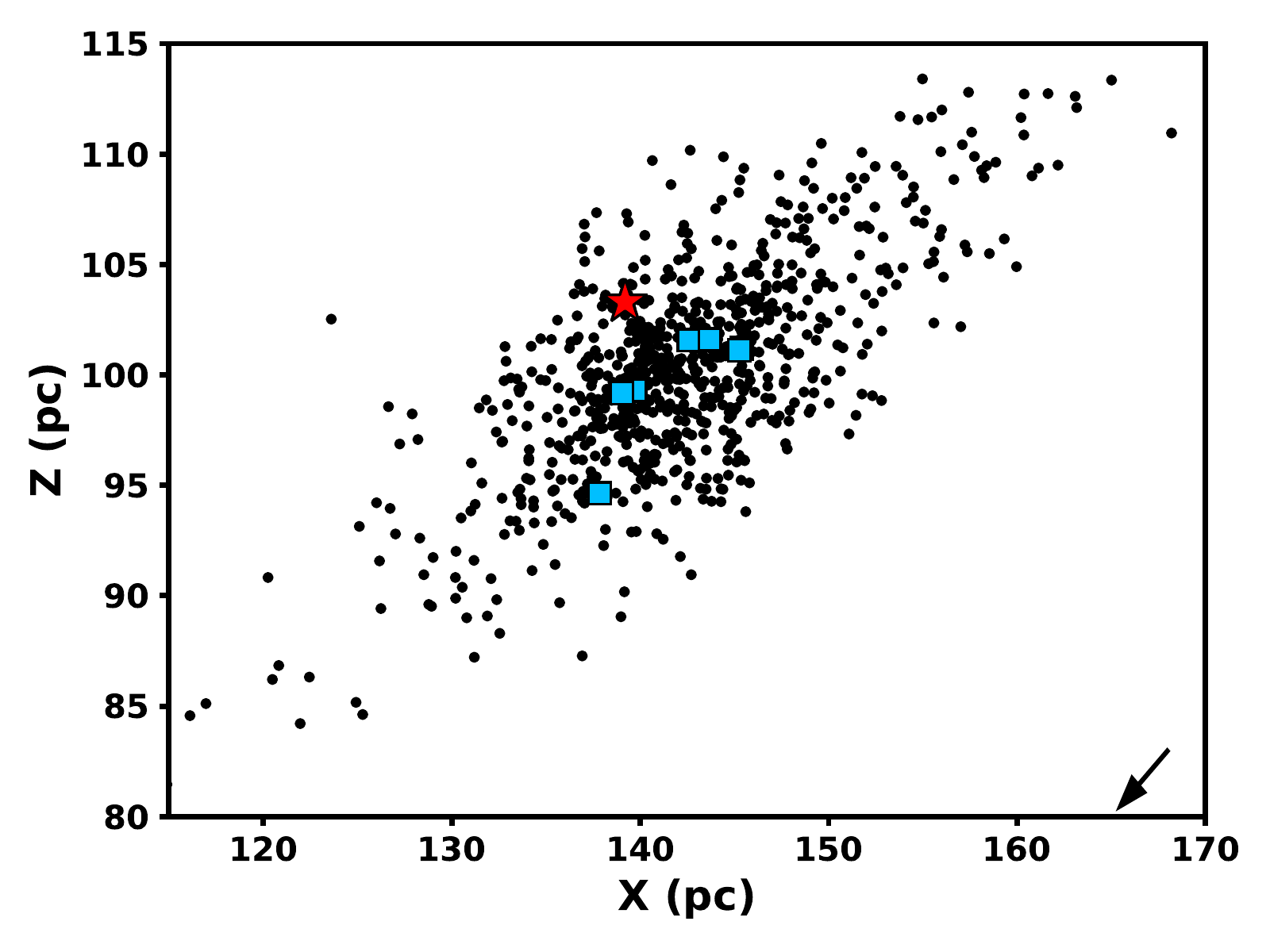}
    \includegraphics[width=0.3\textwidth]{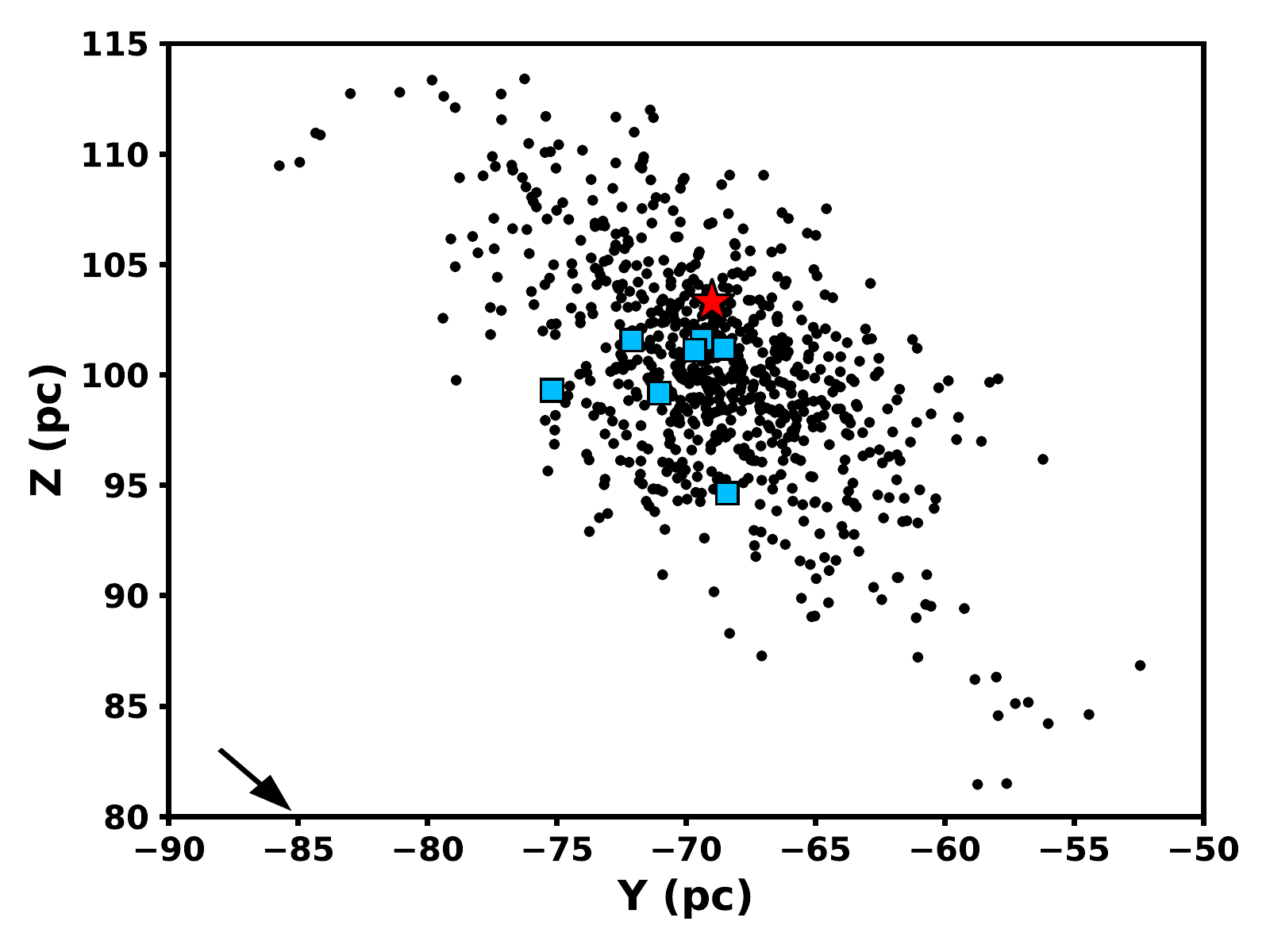}
    \caption{Galactic spatial position of \targ (red star) in relation to other Praesepe members (black points) from \citet{kraushillenbrand_comaber} with membership probability greater than 95\%, computed from \emph{Gaia} DR2 positions and parallaxes. The arrow indicates the typical size and direction of the uncertainty in the positions, which is dominated by the \emph{Gaia} parallax uncertainty. The majority of the objects in front of and behind the core of Praesepe in the line of sight are likely interlopers in the \citet{kraushillenbrand_comaber} membership, introduced due to the lack of parallaxes at the time of selection. \targ sits on the periphery of the central core of the Praesepe cluster. The blue squares indicate the positions of the host stars of the other seven transiting planets in Praesepe identified in \emph{K2} C5 \citep{zeit4}. \label{fig:xyzmap}}
\end{figure*}

\emph{Membership in the Praesepe cluster:} The kinematics, position, and photometry of \targ\ all place it as a high confidence member of the Praesepe cluster. Combining our RV measurement for \targ with the Gaia data release 2 position, proper motions, and parallax measurements allow calculation of the three dimensional space velocity to be $(U, V, W)=(37.3\pm4.6,-18.0\pm2.3,-14.7\pm3.4)$\,km/s. This agrees with the 3D space velocity of Praesepe derived from the locus of the known members updated with \emph{Gaia} DR2 astrometry of  (42,-20,-10)\,km/s with intra-cluster dispersion of 1-2\,km/s \citep{kraushillenbrand_comaber}. Figure \ref{fig:pmproj} shows the proper motion offset from the Praesepe velocity projected onto the plane of the sky for \targ and the Praesepe members of \citet{kraushillenbrand_comaber} with membership probability greater than 95\%.  Here we take the members from \citet{kraushillenbrand_comaber}, but plot the de-projected proper motions from \emph{Gaia} DR2. \targ falls within the range of the velocity dispersion of the members. The \emph{Gaia} DR2 positions and parallax ($\pi = 5.36 \pm 0.06$\,mas; $D = 186.6 ^{+2.1}_{-4.1}$\,pc) place \targ on the periphery of the central core of the Praesepe cluster. Figure \ref{fig:xyzmap} shows the spatial position slices of  known Praesepe members and the position of \targ in relation to the cluster. We calculate a kinematic and spatial membership probability of $\sim$97\% for \targ using the Bayesian membership selection method of \citet{myfirstpaper,wifes1_2015}.

In Figure \ref{fig:prot_praesepe} we also show the \emph{Gaia} (\emph{BP-RP}, \emph{G}) color-magnitude diagram of Praesepe members from \citet{kraushillenbrand_comaber}. The single and binary star sequences are clearly visible, and \targ falls directly on the single star sequence of Praesepe members. In combination with the kinematic and rotational match to the cluster population, this makes membership in Praesepe highly likely. In addition, the narrow single stars sequence rules-out an unresolved companion to \targ  contributing more that 10-20\% of the total observed flux. This is consistent with the lack of companions with $\Delta G\lesssim2$\,mag determined from the \emph{Gaia} astrometry in Section \ref{gaiaastro}.


\emph{Metallicity:} Given the strong membership of \targ in the Praesepe cluster, we can assign the bulk cluster metallicity of the Praesepe population to it. A value of $\mathrm{[Fe/H]}=0.12$ \citep{boesgaard13} is used when required for other calculations and model fitting.

\section{Limits on Additional Planets}\label{limitsonadditionalplanets}
We tested the sensitivity of the combination of our transit search and detrending pipeline and the \emph{K2} data for \targ using the method described in \citet{zeit5}. We injected a series of synthetic planet signals with random parameters into the raw \emph{K2} photometry using the BATMAN model of \citet{batman}. We used orbital periods of 1-30\,days and planet radii of 0.5-10\,R$_{\oplus}$, and allow orbital phase and impact parameter to have values within the interval (0,1). We fixed the eccentricity to zero in these simulations, as it does not significantly alter detectability of a transit, but would significantly increase the required number of trials to obtain a dense enough mapping of parameter space. We randomly injected 5000 trial planet signals for this test. More information regarding this process can be found in \citet{zeit5}.

\begin{figure}
    \centering
    \includegraphics[width=0.5\textwidth]{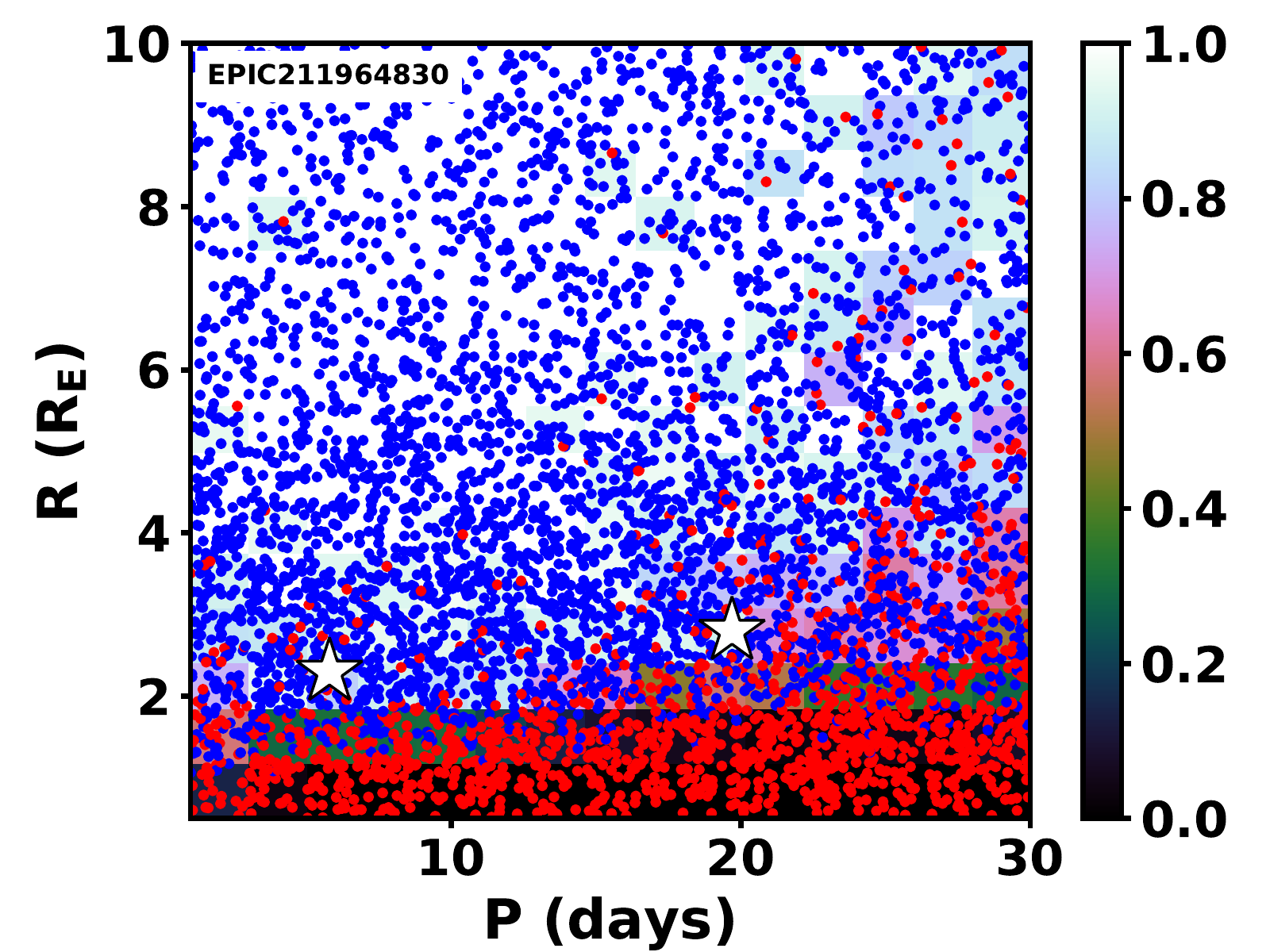}
    \caption{Completeness map for additional planets in the \targ system, produced from injection-recovery testing of our search pipeline. Each point represents an injected planet signal, with blue points indicating recovery and red points non-recovery. White stars are the two detected planets b and c. Our pipeline and the \emph{K2} data for \targ are sensitive to planets as small as $\sim$1-2\,R$_\oplus$ at orbital periods of 1-20\,days.}
    \label{fig:injrec}
\end{figure}

For each injected planet, we applied the corrections for the \emph{K2} pointing and stellar variability, and searched for planets using the BLS algorithm, retaining signals with power spectrum peaks of $>$7-$\sigma$. If a planet was detected within 1\% of both the injected period and injected orbital phase, we flagged it as recovered. Figure \ref{fig:injrec} displays the results of the injection-recovery testing. We found that the combination of the \emph{K2} data and our search methodology is sensitive to 1.7\,R$_{\oplus}$ planets at orbital periods of 1-10\,days, 2.0\,R$_\oplus$ planets at orbital periods of 10-20\,days, and 3.4\,R$_\oplus$ planets out to periods of 25\,days at the 90\% recovery level.

\section{Transit Fitting}\label{transitfitting}
To determine transit parameters for \targtight, we fit the cleaned and detrended \emph{K2} lightcurve with a Markov Chain Monte Carlo (MCMC) as described in \citet{zeit1,zeit6} and \citet{johnson17}. In summary, our MCMC fitting is based on the combination of the BATMAN transit model code \citep{batman} with the Affine-invariant MCMC code \texttt{emcee} \citep{emcee}. The BATMAN transit models were computed including oversampling and binning to the $\sim$30\,minute \emph{K2} long-cadence exposures. We implemented a quadratic limb-darkening law, and used the triangular sampling method of \citet{kipping13}. The free parameters in our model that are different for each planet are the planet-star radius ratio ($R_P/R_*$), orbital period (P), epoch of first transit mid-point (T$_0$), impact parameter (b), and two parameters used in place of eccentricity and argument of periastron ($\sqrt{e}\sin{\omega}$, $\sqrt{e}\cos{\omega}$). These parameters were all fit simultaneously with a common stellar density ($\rho_*$) and the limb darkening parameters (q$_1$ and q$_2$). 

\begin{figure}
    \centering
    \includegraphics[width=0.46\textwidth]{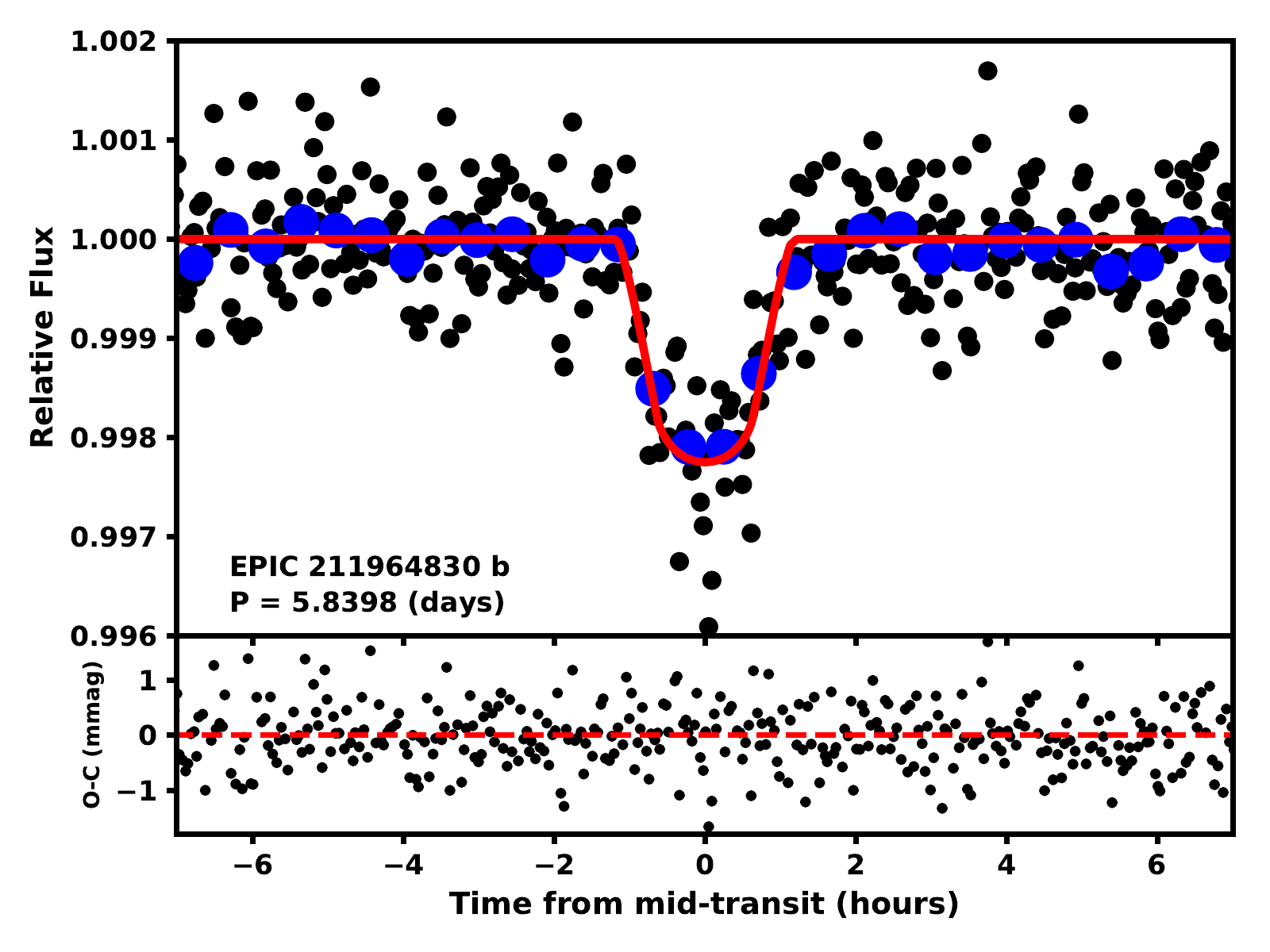}
    \includegraphics[width=0.46\textwidth]{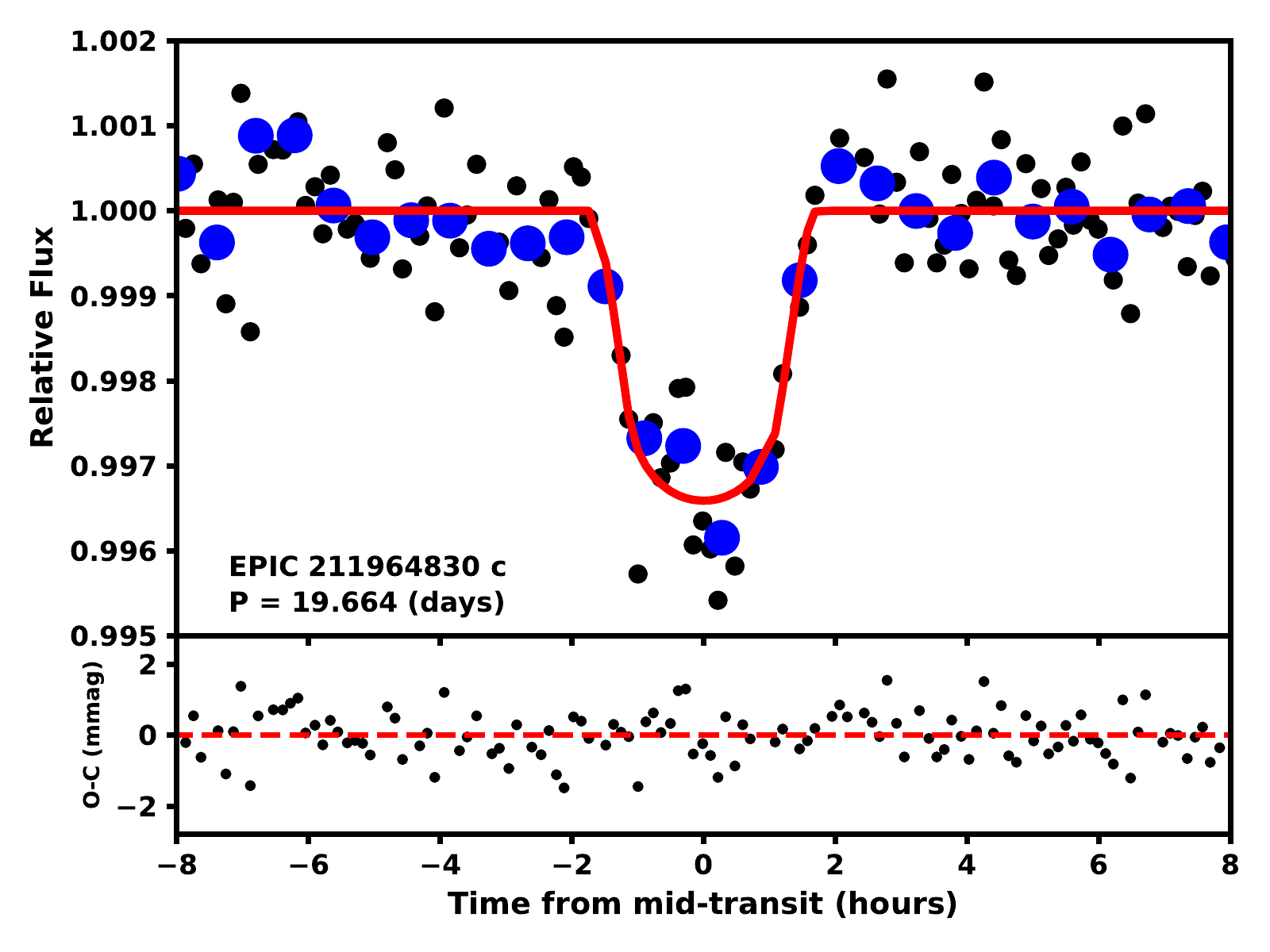}
    \caption{Phase-folded lightcurves centered on the transits for the two detected planets, with best-fit models from our MCMC transit fitting. The black points are the \emph{K2} observations, the blue circles are the binned data, and the red line is the best fit transit model generated with BaTMAN \citep{batman}. The lower panel in each figure shows the best-fit residuals.}
    \label{fig:transmodel}
\end{figure}

We applied a Gaussian prior on stellar density $\rho_*$, determined from our SED fitting described in Section \ref{sec:stelpars}. We also applied a Gaussian prior of the limb darkening parameters determined from the Limb Darkening Tool Kit (LDTK; \citealt{LDTK}) using the \citet{husser13} models, the \emph{Kepler} filter response function, and the stellar parameters from Section \ref{sec:stelpars}. The priors computed were 0.42$\pm$0.10 and 0.38$\pm$0.05 for $u_1$ and $u_2$ respectively. The Gaussian prior was applied after conversion to the triangular sampling parameterization for quadratic limb darkening of \citet{kipping13}. All other parameters were explored with uniform priors with physical boundaries (e.g., $0<b<1$). We ran the MCMC chain for 200,000 steps, with 50,000 steps of burn-in. 

\begin{figure*}
    \includegraphics[width=0.46\textwidth]{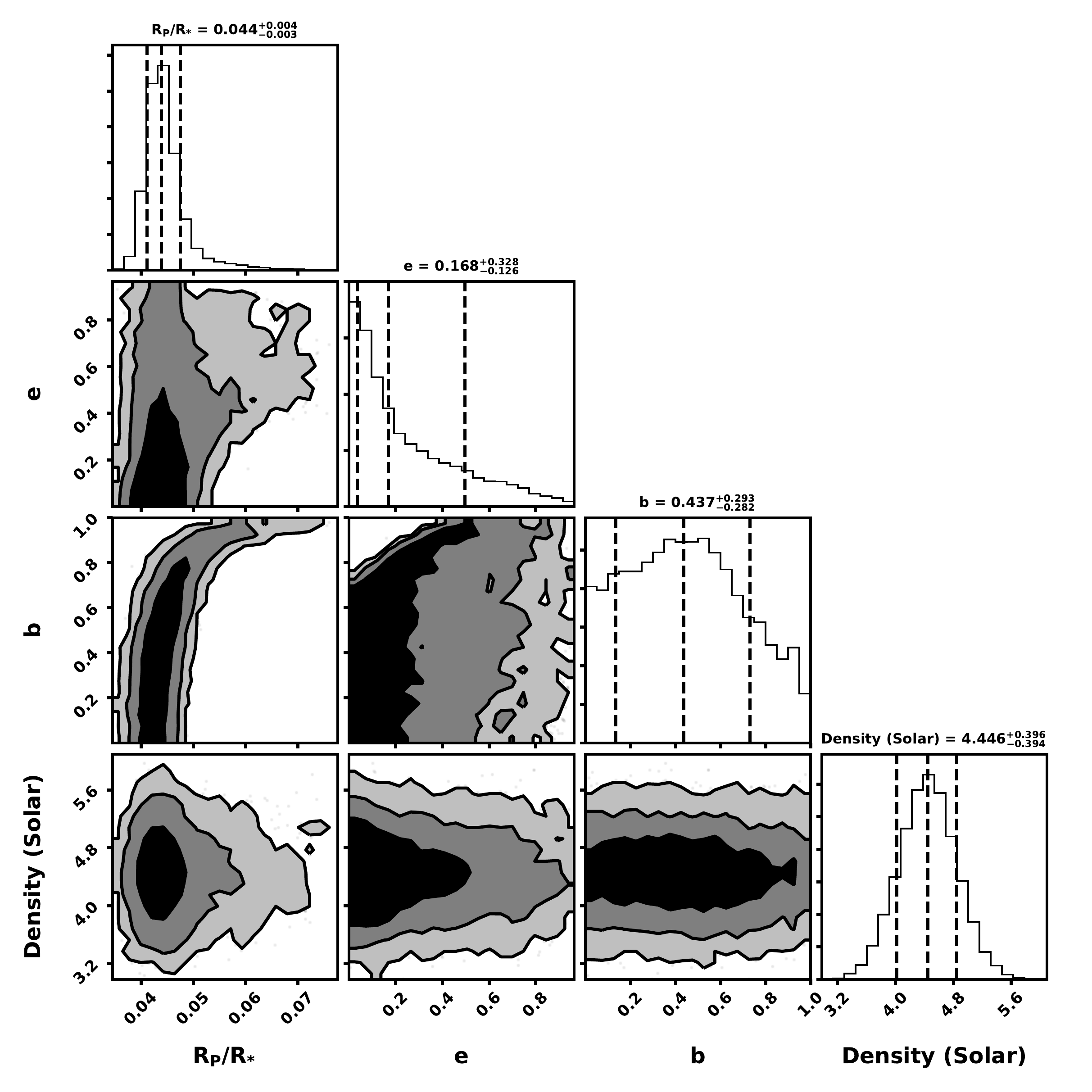}
    \includegraphics[width=0.46\textwidth]{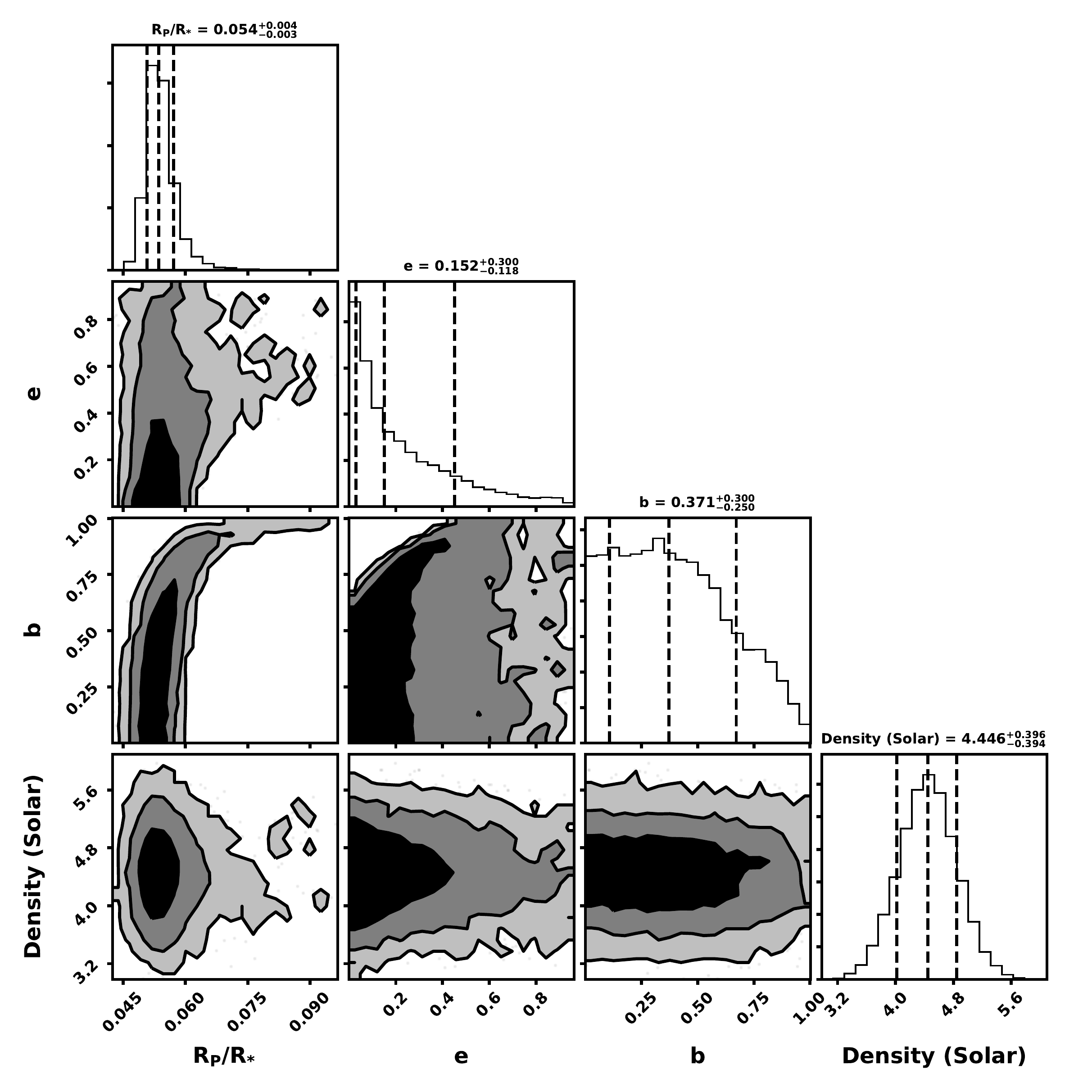}
    \caption{Posteriors for a subset of transit parameters from our MCMC fitting for the inner (left) and outer (right) planets orbiting \targtight. The shaded regions show the 68\%, 95\% and 99.7\% confidence regions, and the dashed lines show the 16$^{th}$, 50$^{th}$ and 84$^{th}$ percentiles respectively.}
    \label{fig:paramcorner}
\end{figure*}

The transit fit parameters and other derived quantities are reported in Table \ref{tab:transfit}. For each value, we report the median, with errors derived from the 16$^{th}$ and 84$^{th}$ percentile values from our fit posteriors. The best fit transit models are shown in Figure \ref{fig:transmodel}. We also show posterior distributions for a subset  of parameters ($R_p/R_*$, $e$, $b$, $\rho_*$) in Figure \ref{fig:paramcorner}. 

Both planets have most likely eccentricities close to zero, which is expected for multiple systems of short-period planets, though both the eccentricity and impact parameters for the planets are not well-constrained by the \emph{K2} data. Both planets are also similar in size with radii of 2.27$^{+0.20}_{-0.16}$\,R$_\oplus$ and 2.77$^{+0.20}_{-0.18}$\,R$_\oplus$.

\section{False Positive Probability}\label{fpp}

While most planet candidates detected by \emph{Kepler} and \emph{K2} are likely to be \textit{bona fide} exoplanets, some transit-like signals may be caused by other astrophysical scenarios. We quantified the likelihood of one of these scenarios causing either of the two transit signals we see towards \targ using the open-source \vespa\ software package \citep{Morton2015}. \texttt{Vespa} calculates the false positive probability (FPP) of transiting signals using the procedure described by \citet{Morton2012} and \citet{Morton2016}. In particular, \vespa\ performs a Bayesian model comparison between several different scenarios which might cause transit-like signals (transiting planets, an eclipsing binary on the target star, an eclipsing binary on a physically bound companion star, or an eclipsing binary on an unassociated background star), and using the transit light curve, stellar parameters, photometric measurements, and observational constraints, determines the likelihood of each model. 

In the case of \targtight, we ran \vespa\ using the transit light curve, broadband photometric measurements from the 2MASS survey\footnote{Previous studies have found that \vespa\ produces more reliable results when the broadband photometry used in its false positive probability calculations only comes from one photometric survey \citep[such as the Kepler Input Catalog or the 2MASS survey,][]{Shporer2017, Mayo2018}.}, and constraints on the presence of nearby stars from the 2MASS \emph{J}-band image of \targtight\footnote{We calculated a ``contrast curve'' by fitting a simple (Moffat function) PSF model to the image of \targtight, subtracting the PSF model away, and calculating the 3$\sigma$ upper limit on the brightness of stars in the residual image.}. Based on these inputs, \vespa\ calculated an FPP of $4\times10^{-3}$ for \targtight\,b and $9\times10^{-4}$ for \targtight\,c. These FPPs do not take into account the fact that candidates in multi-candidate systems are considerably less likely to be false positives than candidates in single-candidate systems \citep{Latham2011, Lissauer2012}. We take this into account by applying a ``multiplicity boost'' to the calculated FPPs for \targtight\,b and c. Following \citet{Lissauer2012}, we divide the calculated FPP for \targtight\,b and c by a factor of 25 as \targ is a two-candidate system. This agrees with the value derived by \citet{sinukoff16} for \emph{K2} data. After applying the multiplicity boost, we find FPPs for \targtight\,b and c of about $10^{-4}$ and $4\times10^{-5}$, respectively. Based on these very low FPPs, we consider both candidates in the \targ system to be validated planets. 

\section{Discussion}\label{discussion}
We have reported the discovery and characterization of a two planet transiting system in the Praesepe open cluster. There are now several detected transiting planets in young open clusters and associations observed by \emph{K2}, though \targ is one of only two multiple-planet systems, the other being K2-136, a three transiting-planet system in the Hyades open cluster \citep{zeit6}.

\begin{figure}
    \centering
    \includegraphics[width=0.49\textwidth]{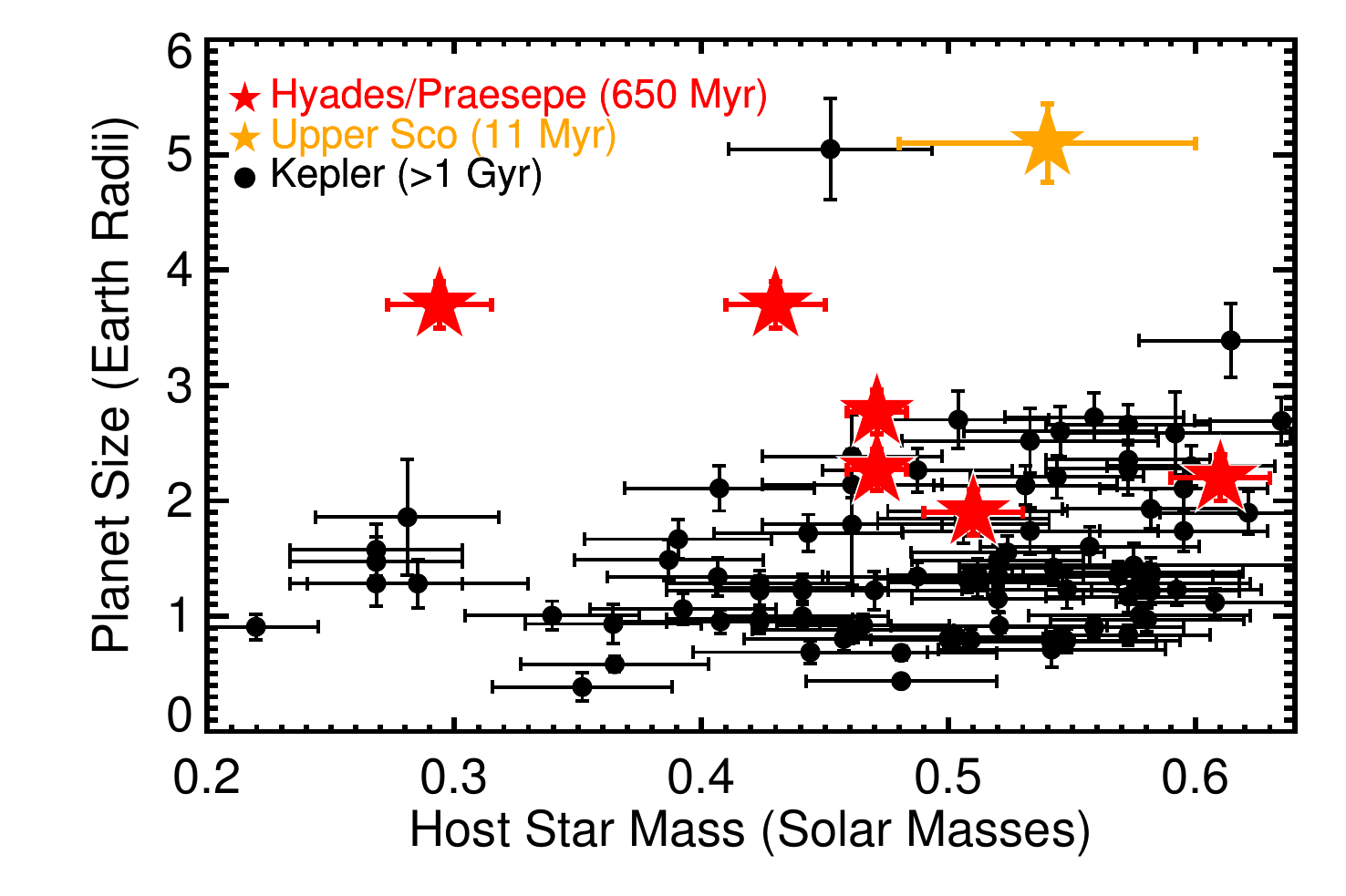}
    \caption{Host star mass and planet radii for the seven transiting planetary systems in Praesepe and the Hyades from \emph{K2} C4/5 \citep{zeit1,zeit4,zeit6} and those presented in this paper from C16, compared to older M-dwarf hosted planets from the original \emph{Kepler} samples \citep{dressing15}. The 650\,Myr Praesepe and Hyades planet population have larger radii than those hosted by older M dwarfs. The single 10\,Myr old planet in Upper Scorpius (K2-33\,b; \citealt{zeit3}) is also significantly larger than the older planets.}
    \label{fig:popradii}
\end{figure}

\targtight\,b and c are both likely mini-Neptunes, and both sit near the upper envelope of the field mass--panet radius distribution, as is seen for other planets in intermediate-age clusters. These  two planets continue the trend of young open clusters M dwarfs hosting planets of larger radii than have been observed for planets transiting older field population dwarfs from the original \emph{Kepler} sample \citep{zeit4,dressing15}. Figure \ref{fig:popradii} shows the planet radii and host star masses of the M-dwarf hosted young planets identified in the ZEIT survey \citep{zeit1,zeit3,zeit4,zeit6}, including \targtight\,b/c, compared to older transiting systems. The possible inflation in radii at $\sim$650\,Myr may be a sign of ongoing atmosphere loss (e.g., \citealt{Lopez2012}). With further completeness testing on the entire sample of Hyades and Praesepe stars observed by \emph{K2} a measure of the rate and significance of the potential radius difference could be measured.

Systems with multiple transiting planets offer the potential for many science cases not possible with single transit systems. In particular, eccentricity and stellar density can strongly constraint each other \citep{vaneylen15,mann17a}. Planet masses for multiple systems can also be measured from transit timing variations (TTV's) \citep{hadden17}. Though we did not explicitly test for TTV's, the detection of TTV's in the \emph{K2} dataset is unlikely; similar size planets show variations of $<$15\,min, which is smaller than the long-cadence timing of $\simeq$30\,mins. In particular, TTV's on planet b due to planet c are expected to be very small ($<$1\,min) given that the orbital periods are very far from a resonance \citep{agol05}. One scenario where TTV detection could be possible involves the presence of a third planet in or near e.g., a 2:1 resonance with the inner planet b. Such a planet would have to be approximately earth-mass to have avoided detection in the \emph{K2} lightcurve. The TTV amplitude from such a planet on the ephemeris of \targtight\,b, assuming zero eccentricities, is 5-15\,min depending on the proximity to resonance \citep{agol05}.

The currently available long-cadence data from \emph{K2} is particularly unsuited to the science cases described above. However, \targ is highly amenable to follow-up photometry. Both planets are large enough that ground-based facilities could resolve their transits ($\simeq$3\,mmag), though the faintness of the host star ($r\simeq 16$mag) may be prohibitive for small apertures at high cadence.  Shorter cadence data resolving ingress and egress shapes can place stronger constraints on eccentricity, and offer suggestions as to the types of formation mechanisms responsible for forming these two short-period planets. Space-based follow-up with the Hubble Space Telescope or Spitzer is possible for both planets. In Spitzer channel 1 ($\simeq$3.5\,$\mu m$; \citealt{hora08}) \targ is $\simeq$12\,mag \citep{wise10} and in a 2\,min exposure a SNR of 500\,pmm is possible. This is sufficient to resolve the transit shape from even a single transit.

Follow-up spectroscopy to measure the masses of \targtight\,b,c may not be possible. \targ shows stellar variability with a period 22.8\,days and photometric amplitude of $\simeq$3\%. If the star is seen equator-on, this amplitude of variability is expected to produce RV variability of $\simeq$30\,m/s in a similar band as \emph{K2}. Using the mass-radius relation for planets from \citet{weiss14} and the radii inferred from our transit fitting, we find that \targtight\,b,c have likely masses of 5.8\,M$_\oplus$ and 7\,M$_\oplus$ respectively. Assuming circular orbits and the stellar properties derived above, these masses correspond to radial velocity semi-amplitudes of 3.4\,m/s and 2.7\,m/s respectively. The amplitude of these signals is significantly smaller than the expected stellar rotations signal. Moving to the near-infrared, where the stellar variability is expected to have significantly smaller amplitude, could alleviate this problem in combination with our prior knowledge of the rotation period of the star.

\section*{Acknowledgments}
ACR was supported as a 51 Pegasi b Fellow though the Heising-Simons Foundation. AWM was supported through NASA Hubble Fellowship grant 51364 awarded by the Space Telescope Science Institute, which is operated by the Association of Universities for Research in Astronomy, Inc., for NASA, under contract NAS 5-26555. AV's work was performed under contract with the California Institute of Technology (Caltech)/Jet Propulsion Laboratory (JPL) funded by NASA through the Sagan Fellowship Program executed by the NASA Exoplanet Science Institute. S.T.D.~acknowledges support provided by the NSF through grant AST-1701468. This paper includes data collected by the K2 mission. Funding for the K2 mission is provided by the NASA Science Mission directorate. Some of the data presented in this paper were obtained from the Mikulski Archive for Space Telescopes (MAST). STScI is operated by the Association of Universities for Research in Astronomy, Inc., under NASA contract NAS5-26555. The authors acknowledge the Texas Advanced Computing Center (TACC) at The University of Texas at Austin for providing HPC resources that have contributed to the research results reported within this paper\footnote{http://www.tacc.utexas.edu}.  This work has made use of data from the European Space Agency (ESA) mission \emph{Gaia} \footnote{https://www.cosmos.esa.int/gaia}, processed by the \emph{Gaia} Data Processing and Analysis Consortium (DPAC)\footnote{https://www.cosmos.esa.int/web/gaia/dpac/consortium}. Funding for the DPAC has been provided by national institutions, in particular the institutions participating in the \emph{Gaia} Multilateral Agreement. This research has made use of the VizieR catalogue access tool, CDS, Strasbourg, France. The original description of the VizieR service was published in A\&AS 143, 23. This research has made use of NASA's Astrophysics Data System Bibliographic Services.

\facilities{Kepler, IRTF , Texas Advanced Computing Center}

\bibliographystyle{apj}
\bibliography{ms.bbl}
\clearpage

\begin{deluxetable*}{ccc}
\tabletypesize{\scriptsize}
\tablewidth{0pt}
\tablecaption{Properties of the host star \targ (EPIC 211964830). \label{proptab}}
\tablehead{\colhead{Parameter} & \colhead{Value} & \colhead{Source} }
\startdata
\multicolumn{3}{c}{Astrometry}\\
\hline
$\alpha$ R.\,A.  & 08 45 26.054 & EPIC\\
$\delta$ Decl. & +19 41 54.46 & EPIC\\
$\mu_\alpha$ (mas\,yr$^{-1}$)& -37.900$\pm$0.095 & \emph{Gaia} DR2\\
$\mu_\delta$  (mas\,yr$^{-1}$) & -13.079$\pm$0.061 & \emph{Gaia} DR2\\
$\pi$ (mas) & 5.3598$\pm$0.0605 & \emph{Gaia} DR2\\
\hline
\multicolumn{3}{c}{Photometry}\\
\hline
G$_{Gaia}$ (mag) & 15.6625$\pm$0.0006 & \emph{Gaia} DR2\\
BP$_{Gaia}$ (mag) & 16.9463$\pm$0.006 & \emph{Gaia} DR2\\
RP$_{Gaia}$ (mag) &  14.5382$\pm$0.0015 & \emph{Gaia} DR2\\
g (mag) & 17.259$\pm$0.006  & PanSTARRS\\
r (mag) &16.075$\pm$0.002  & PanSTARRS\\
i (mag) & 14.965$\pm$0.003 & PanSTARRS\\
z (mag) &  14.471$\pm$0.002 & PanSTARRS\\
r (mag) & 16.052$\pm$0.031 & CMC15\\
J (mag) &13.047$\pm$0.025& 2MASS\\
H (mag) &12.386$\pm$0.022& 2MASS\\	
Ks (mag) &12.183$\pm$0.020& 2MASS\\
W1 (mag) & 12.048$\pm$0.023 & WISE\\
W2 (mag)& 11.978$\pm$0.023 & WISE\\
\hline
\multicolumn{3}{c}{Kinematics \& Position}\\
\hline
Barycentric RV (km\, s$^{-1}$) & 26$\pm$6 & This paper\\
U (km\, s$^{-1}$) &37.3$\pm$4.6 & This paper\\
V (km\, s$^{-1}$) & -18.0$\pm$2.6 & This paper\\
W (km\, s$^{-1}$) & -14.7$\pm$3.5 & This paper\\
X (pc) & 139.2$\pm$1.6 & This paper\\
Y (pc) & -69.0$\pm$0.8 & This paper\\
Z (pc) & 103.3$\pm$1.2 & This paper\\
Distance (pc) &186.6$^{+2.1}_{-4.1}$& \emph{Gaia} DR2 \\
\hline
\multicolumn{3}{c}{Physical Properties}\\
\hline
Rotation Period (days) &22.8$\pm$0.6 & This paper\\
Spectral Type  & M2.5$\pm$0.5& This paper \\
\fbol\,(erg\,cm$^{-2}$\,s$^{-1}$)& $3.068\pm0.068\times10^{-11}$ & This paper\\
T$_{\mathrm{eff}}$ (K) &3580$\pm$70& This paper\\
M$_\star$ (M$_\odot$) & 0.471$\pm$0.012 & This paper \\
R$_\star$ (R$_\odot$) &  0.473$\pm$0.014 & This paper \\
L$_\star$ (L$_\odot$) & 0.0330$\pm$0.0012 & This paper \\
$\rho_\star$ ($\rho_\odot$) & 4.5$\pm$0.4 & This paper \\
$[$Fe/H$]$ &0.12$\pm$0.04 & Praesepe \citep{boesgaard13} \\
\enddata
\end{deluxetable*}

\begin{deluxetable*}{ccc}
\tabletypesize{\scriptsize}
\tablewidth{0pt}
\tablecaption{Transit Fit Parameters. \label{tab:transfit}}
\tablehead{\colhead{Parameter} & \colhead{Planet\,b} & \colhead{Planet\,c}}
\startdata
Period (days)               & 5.839770$^{+0.000061}_{-0.000063}$&  19.663650$^{+0.000303}_{0.000306}$\\
$R_P/R_*$                   & 0.0439$^{+0.0036}_{-0.0026}$ & 0.0536$^{+0.0035}_{-0.0027}$  \\
T$_0$ (BJD-2400000)         & 58102.09356$^{+0.00046}_{-0.00046}$& 58096.93729$^{+0.00077}_{-0.00071}$  \\
Impact Parameter            & 0.44$^{+0.29}_{-0.28}$ & 0.37$^{+0.30}_{-0.25}$\\
Duration$^a$ (hours)        &1.88$_{-0.39}+{+0.17}$& 2.92$_{-0.50}+{+0.20}$\\
Inclination$^a$ (degrees)   &88.9$^{+0.7}_{-0.7}$& 89.6$^{+0.3}_{-0.3}$   \\
$a/R_*^a$                     &22.4$^{+0.7}_{-0.7}$&50.4$^{+1.5}_{-1.6}$   \\
Eccentricity$^b$                &$<$0.50 &$<$0.45\\
$R_P^c$ (R$_\oplus$)        &2.27$^{+0.20}_{-0.16}$&2.77$^{+0.20}_{-0.18}$   \\
T$_\mathrm{eq}^c$ (K)       &489$^{+12}_{-13}$&326$^{+8}_{-9}$\\
\hline
\multicolumn{3}{c}{Global Parameters}\\
\hline
$\rho_*$ ($\rho_\odot$)     &\multicolumn{2}{c}{4.45$^{+0.39}_{-0.40}$}\\
$u_1$                       &\multicolumn{2}{c}{0.42$^{+0.09}_{-0.09}$}\\
$u_2$                       &\multicolumn{2}{c}{0.27$^{+0.08}_{-0.08}$}\\
\enddata
\tablenotetext{a}{Inclination, $\omega$, $a/R_*$ and transit duration were not fit as part of our MCMC, but were derived from other fit parameters (see Section \ref{transitfitting}). 
\tablenotetext{b}{The most likely eccentricities for both systems is $\sim$0, and so we report only the 1-$\sigma$ upper limit.} \tablenotetext{c}{$R_P$ T$_\mathrm{eq}$ were calculated using T$_\mathrm{eff}$ from Section (\ref{sec:stelpars}). Equilibrium temperature T$_\mathrm{eq}$ was calculated assuming an albedo of 0.3.}}
\end{deluxetable*}

\end{document}